\documentclass[a4paper,12pt]{article}
\pdfoutput=1 
\usepackage{geometry}
\usepackage{amsmath,amssymb,amsfonts}
\usepackage{xcolor,graphicx,cite,soul}
\usepackage{array,booktabs,longtable}
\usepackage{caption,microtype}
\usepackage{hyperref}

\newcommand{\lsim}
{\;\raisebox{-.3em}{$\stackrel{\displaystyle <}{\sim}$}\;}

\newcommand\Code[1]{\ensuremath{\texttt{#1}}}

\newcommand\tb{\tan\beta}
\newcommand\TB{t_\beta}

\newcommand\ReDiag{\mathop{%
  \raise .5pt\hbox{[}%
  \widetilde{\mathrm{Re}}%
  \raise .5pt\hbox{]}}}
\newcommand\ReOffDiag{\mathop{%
  \raise .5pt\hbox{$\llbracket$}%
  \widetilde{\mathrm{Re}}%
  \raise .5pt\hbox{$\rrbracket$}}}

\newcommand\MSbar{\ensuremath{\overline{\mathrm{MS}}}}

\newcommand\cL{{\cal L}}

\newcommand\MHp{M_{H^\pm}}

\newcommand\Ab{A_b}
\newcommand\At{A_t}

\newcommand\Sf{\tilde f}

\newcommand\Sn{\tilde\nu}
\newcommand\Sl{\tilde l}

\newcommand\Se{\mathrm{\tilde e}}

\newcommand\Fe{\mathrm{e}}

\newcommand\Fu{\mathrm{u}}

\newcommand\Fd{\mathrm{d}}

\newcommand\cpri{c^\prime}

\newcommand\npri{n^\prime}

\newcommand\spri{s^{\prime}}

\newcommand\ino[1]{\tilde\chi_{#1}}

\newcommand\chapm[1]{\ino{#1}^\pm}
\newcommand\champ[1]{\ino{#1}^\mp}
\newcommand\chap[1]{\ino{#1}^+}
\newcommand\cham[1]{\ino{#1}^-}
\newcommand\cha{\chapm}
\newcommand\mcha[1]{m_{\chapm{#1}}}

\newcommand\neu[1]{\ino{#1}^0}
\newcommand\mneu[1]{m_{\neu{#1}}}

\newcommand\refta[1]{Tab.~\ref{#1}}
\newcommand\reftas[1]{Tabs.~\ref{#1}}

\newcommand\citere[1]{Ref.~\cite{#1}}
\newcommand\citeres[1]{Refs.~\cite{#1}}

\newcommand\eg{e.g.}
\newcommand\ie{i.e.\ }
\newcommand\wrt{w.r.t.\ }

\newcommand{\CP}{{\cal CP}}
\newcommand{\cp}{{\CP}}

\newcommand{\tev}{\,\, \mathrm{TeV}}
\newcommand{\gev}{\,\, \mathrm{GeV}}
\newcommand{\mev}{\,\, \mathrm{MeV}}

\newcommand{\eecc}{\ensuremath{e^+e^- \to \chapm{c} \champ{\cpri}}}
\newcommand{\eecece}{e^+e^- \to \chap1 \cham1}
\newcommand{\eececz}{e^+e^- \to \chapm1 \champ2}
\newcommand{\eeczcz}{e^+e^- \to \chap2 \cham2}

\newcommand{\eenn}{\ensuremath{e^+e^- \to \neu{n} \neu{\npri}}}
\newcommand{\eenene}{e^+e^- \to \neu1 \neu1}

\newcommand{\eenend}{e^+e^- \to \neu1 \neu3}

\newcommand{\eeSlSl}{\ensuremath{e^+e^- \to \Sl_{gs} \Sl_{gs^{\prime}}}}
\newcommand{\eeSeSe}{\ensuremath{e^+e^- \to \Se^{\pm}_{gs} \Se^{\mp}_{gs^{\prime}}}}

\newcommand{\eeSaeSae}{\ensuremath{e^+e^- \to \tilde{\tau}^+_1 \tilde{\tau}^-_1}}

\newcommand{\eeSnSn}{\ensuremath{e^+e^- \to \Sn_{g} \Sn_{g}}}

\newcommand\FA{\texttt{FeynArts}}
\newcommand\FC{\texttt{FormCalc}}
\newcommand\LT{\texttt{LoopTools}}
\newcommand\FH{\texttt{FeynHiggs}}

\newcommand\fb{\ensuremath{\mbox{fb}}}
\newcommand\ab{\ensuremath{\mbox{ab}}}

\newcommand\iab{\ensuremath{\ab^{-1}}}

\newcommand\msneu{m_{\tilde{\nu}_{e,\mu,\tau}}}

\newcommand{\Sce}{S1}
\newcommand{\Scz}{S2}

\newcommand{\sig}{\sigma}
\newcommand{\sigfull}{\sigma_{\text{full}}}
\newcommand{\sigtree}{\sigma_{\text{tree}}}
\newcommand{\sigloop}{\sigma_{\text{loop}}}

\newcommand{\phiAeg}{\varphi_{A_{\Fe_g}}}

\def\order#1{\ensuremath{{\cal O}(#1)}}
\def\reffi#1{\mbox{Fig.~\ref{#1}}}
\def\reffis#1{\mbox{Figs.~\ref{#1}}}

\def\ga{\gamma}

\def\phiAt{\varphi_{\At}}

\def\phiMe{\varphi_{M_1}}

\def\MSL{M_{\tilde L}}
\def\MSE{M_{\tilde E}}

\definecolor{Orange}{named}{orange}
\definecolor{Purple}{named}{purple}
\definecolor{Lightblue}{cmyk}{0.9,0.1,0.1,0.3}
\definecolor{dgelborange}{cmyk}{0.,0.3,0.5, 0.}
\definecolor{Lila}{rgb}{0.5,0.,1}

\graphicspath{{figs/}}
\allowdisplaybreaks
\begin{document}
\thispagestyle{empty}

\def\thefootnote{\fnsymbol{footnote}}

\begin{flushright}
\mbox{}
IFT--UAM/CSIC--18-006 \\
%arXiv:1704.07627 [hep-ph]
\end{flushright}

\vspace{0.5cm}

\begin{center}

{\large\sc 
{\bf Production of Electroweak SUSY Particles at ILC and CLIC}}

%\vspace{0.4cm}

\vspace{1cm}

{\sc
S.~Heinemeyer$^{1\,2\,3}$%
\footnote{email: Sven.Heinemeyer@cern.ch}%
\footnote{Talk presented at the International Workshop
on Future Linear Colliders (LCWS2017),\\
\mbox{}\hspace{5mm} Strasbourg, France, 23-27 October 2017.}
~and C.~Schappacher$^{4}$%
\footnote{email: schappacher@kabelbw.de}%
}

\vspace*{.7cm}

{\sl
$^1$Campus of International Excellence UAM+CSIC, 
Cantoblanco, 28049, Madrid, Spain 

\vspace*{0.1cm}

$^2$Instituto de F\'isica Te\'orica (UAM/CSIC), 
Universidad Aut\'onoma de Madrid, \\ 
Cantoblanco, 28049, Madrid, Spain

\vspace*{0.1cm}

$^3$Instituto de F\'isica de Cantabria (CSIC-UC), 
39005, Santander, Spain

\vspace*{0.1cm}

$^4$Institut f\"ur Theoretische Physik,
Karlsruhe Institute of Technology, \\
76128, Karlsruhe, Germany (former address)
}

\end{center}

\vspace*{0.1cm}

\begin{abstract}
\noindent
For the search for electroweak particles in the Minimal Supersymmetric 
Standard Model (MSSM) as well as for future precision analyses of these 
particles an accurate knowledge of their production and decay properties
is mandatory. We review the 
evaluation of the cross sections for the chargino, neutralino and scalar
lepton production  
at $e^+e^-$ colliders in the MSSM with complex parameters (cMSSM). 
The evaluation is based on a full one-loop calculation of the various
production mechanisms, including soft and hard photon radiation.  
The dependence of the chargino/neutralino/slepton cross sections on the
relevant  
cMSSM parameters is analyzed numerically.  We find sizable contributions 
to many production cross sections.  They amount roughly $\pm 15\,\%$
of the tree-level results, but can go up to $\pm 40\,\%$ or higher in
extreme cases. 
Also the complex phase dependence of the one-loop corrections was found
non-negligible.  The full one-loop contributions are thus crucial for 
physics analyses at a future linear $e^+e^-$ collider such as the ILC or 
CLIC.
\end{abstract}

%\pacs{}

\def\thefootnote{\arabic{footnote}}
\setcounter{page}{0}
\setcounter{footnote}{0}

\newpage

%%%%%%%%%%%%%%%%%%%%%%%%%%%%%%%%%%%%%%%%%%%%%%%%%%%%%%%%%%%%%%%%%%%%%%%%%%%%%%%
%%%%%%%%%%%%%%%%%%%%%%%%%%%%%%%%%%%%%%%%%%%%%%%%%%%%%%%%%%%%%%%%%%%%%%%%%%%%%%%

\section{Introduction}
\label{sec:intro}

One of the most important tasks at the LHC is to search for physics beyond the 
Standard Model (SM), where the Minimal Supersymmetric Standard Model 
(MSSM)~\cite{mssm,HaK85,GuH86} is one of the leading candidates.
Supersymmetry (SUSY) predicts two scalar partners for all SM fermions as well
as fermionic partners to all SM bosons. 
In particular two scalar quarks or scalar leptons are predicted for each
SM quark or lepton. Concerning the Higgs-boson sector, 
contrary to the case of the SM, in the MSSM two Higgs doublets are required.
This results in five physical Higgs bosons instead of the single Higgs
boson in the SM. These are the light and heavy $\cp$-even Higgs bosons, $h$
and $H$, the $\cp$-odd Higgs boson, $A$, and the charged Higgs bosons,
$H^\pm$. At tree-level the Higgs sector is described by the mass of the
charged Higgs boson, $\MHp$ and the ratio of the two vacuum expectation
values, $\tb \equiv \TB := v_2/v_1$. Higher-order corrections are
crucial to yield reliable predictions in the MSSM Higgs-boson sector, see
\citeres{habilSH,awb2,PomssmRep} for reviews.
The neutral SUSY partners of the (neutral) Higgs and electroweak gauge
bosons are the four neutralinos, $\neu{1,2,3,4}$. The corresponding
charged SUSY partners are the charginos, $\cha{1,2}$.

If SUSY is realized in nature and the scalar quarks and/or the gluino
are in the kinematic reach of the LHC, it is expected that these
strongly interacting particles are copiously produced eventually. 
On the other hand, SUSY particles that interact only via the electroweak
force, \ie\ the charginos, neutralinos and sleptons, have a much smaller
production cross section at the LHC. Correspondingly, the LHC
discovery potential as well as the current experimental bounds are
substantially weaker.

At a (future) $e^+e^-$ collider charginos, neutralinos and sleptons,
depending on their masses and the available center-of-mass energy, could
be produced and analyzed in detail, see \eg\ \citere{lcws-MC-proc}. 
Corresponding studies can be found for the ILC in
\citeres{ILC-TDR,teslatdr,ilc,LCreport} and for CLIC in
\citeres{CLIC,LCreport}. 
(Results on the combination of LHC and ILC results can be found in 
\citere{lhcilc}.) Such precision studies will be crucial to determine
the nature of those particles and the underlying SUSY parameters.

In order to yield a sufficient accuracy, one-loop corrections to the 
various chargino/neutralino/slepton production and decay modes have to
be considered. 
Full one-loop calculations in the cMSSM for various
chargino/neutralino/slepton decays in the cMSSM have been presented over
the last years~\cite{Stau2decay,LHCxC,LHCxN,LHCxNprod}.
One-loop corrections for their production from the decay of Higgs bosons
(at the LHC or ILC/CLIC) can be found in
\citeres{HiggsDecaySferm,HiggsDecayIno}.  
Here we review the recent calculations of chargino/neutralino 
production at $e^+e^-$ colliders~\cite{eeIno} and give a preview of the
slepton production at $e^+e^-$ colliders~\cite{eeSlep}. The following
channels are considered:
\begin{align}
\label{eq:eecc}
&\sig(\eecc) \qquad (c,\cpri = 1,2)\,, \\
\label{eq:eenn}
&\sig(\eenn) \qquad (n,\npri = 1,2,3,4)\,, \\
\label{eq:eeSeSe}
&\sig(\eeSeSe) \qquad (s,\spri = 1,2)\,, \\
\label{eq:eeSnSn}
&\sig(\eeSnSn)\,,
\end{align}
with $\Se_g = \{\tilde e, \tilde\mu, \tilde\tau\}$, 
$\Sn_g = \{\Sn_e, \Sn_\mu, \Sn_\tau\}$, and the generation index $g = 1,2,3$.
Our evaluation of the four channels (\ref{eq:eecc}) -- (\ref{eq:eeSnSn}) 
is based on a full one-loop calculation, \ie including electroweak (EW) 
corrections, as well as soft and hard QED radiation. 
The renormalization scheme employed is the same one as for the decay 
of charginos/neutralinos/sleptons~\cite{Stau2decay,LHCxC,LHCxN,LHCxNprod}.
Consequently, the predictions for the production and decay can be 
used together in a consistent manner.

%%%%%%%%%%%%%%%%%%%%%%%%%%%%%%%%%%%%%%%%%%%%%%%%%%%%%%%%%%%%%%%%%%%%%%%%%%%%%%%
%%%%%%%%%%%%%%%%%%%%%%%%%%%%%%%%%%%%%%%%%%%%%%%%%%%%%%%%%%%%%%%%%%%%%%%%%%%%%%%

\section{The production processes at one-loop}
\label{sec:calc}

Here we briefly review the contributing loop diagrams to the processes
(\ref{eq:eecc}) -- (\ref{eq:eeSnSn}). 
The diagrams and corresponding amplitudes have been obtained 
with \FA~\cite{feynarts}, using the MSSM model file 
(including the MSSM counterterms) of \citere{MSSMCT}. 
The further evaluation has been performed with \FC\ and \LT~\cite{formcalc}. 
The specific versions of the codes used can be found in
\citeres{eeIno,eeSlep}. All relevant details about the various sectors
of the cMSSM and their renormalization as well as on the cancellation of
the UV, IR and collinear divergences can also be found in
\citeres{eeIno,eeSlep}, see also the descriptions given in 
\citeres{HiggsDecaySferm,HiggsDecayIno,MSSMCT,SbotRen,Stop2decay,%
Gluinodecay,Stau2decay,LHCxC,LHCxN,LHCxNprod,HiggsProd,HpProd}.

%%%%%%%%%%%%%%%%%%%%%%%%%%%%%%%%%%%%%%%%%%%%%%%%%%%%%%%%%%%%%%%%%%%%%%%%%%%%%%%

%%%%%%%%%%%%%%%%%%%%%%%%%%%%%%%%%%%%%%%%%%%%%%%%%%%%%%%%%%%%%%%%%%%%%%%%%%%%%%%

\subsection{Contributing diagrams for chargino/neutralino production}
\label{sec:diagrams-chaneu}

%Fig1
%%%%%%%%%%%%%%%%%%%%%%%%% F I G U R E %%%%%%%%%%%%%%%%%%%%%%%%%%%%%%%%%%%%%%%%%
\begin{figure}
\begin{center}
\framebox[14cm]{\includegraphics[width=0.21\textwidth]{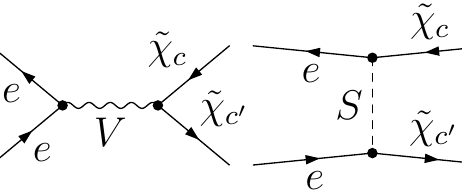}}
\framebox[14cm]{\includegraphics[width=0.67\textwidth]{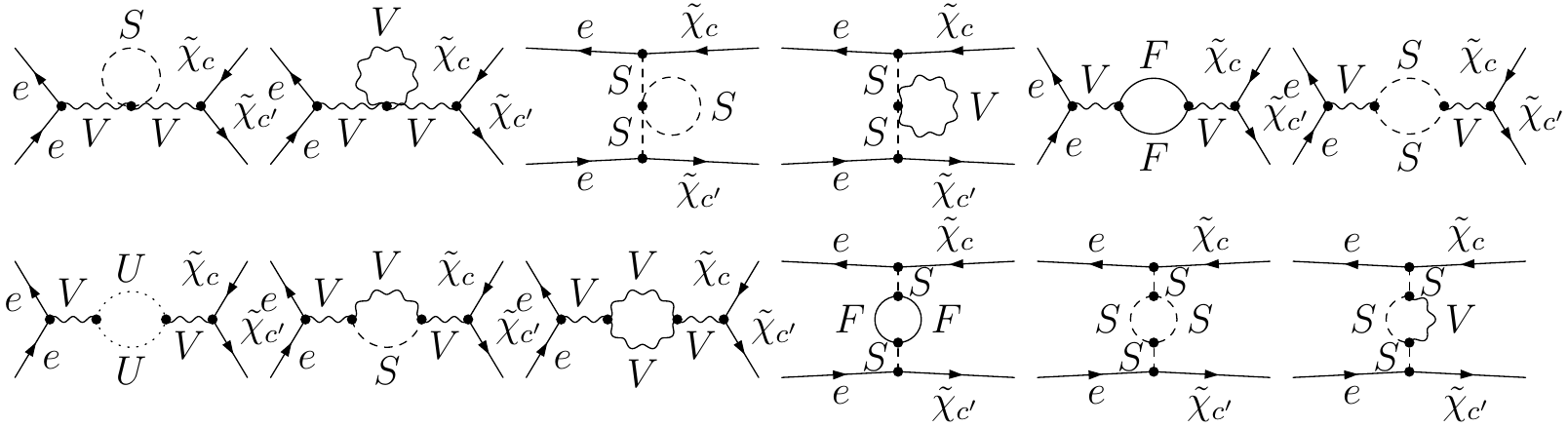}}
\framebox[14cm]{\includegraphics[width=0.67\textwidth]{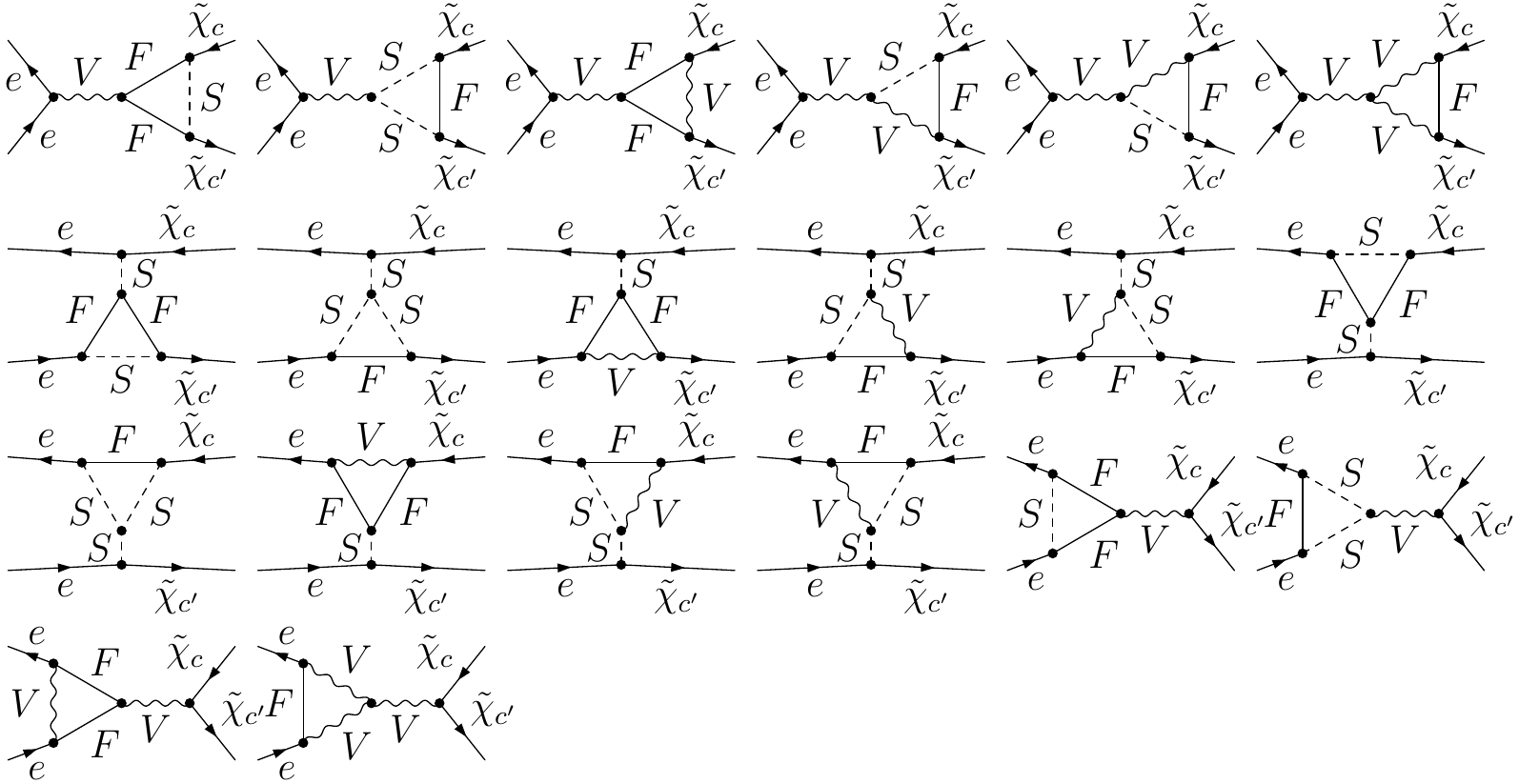}}
\framebox[14cm]{\includegraphics[width=0.67\textwidth]{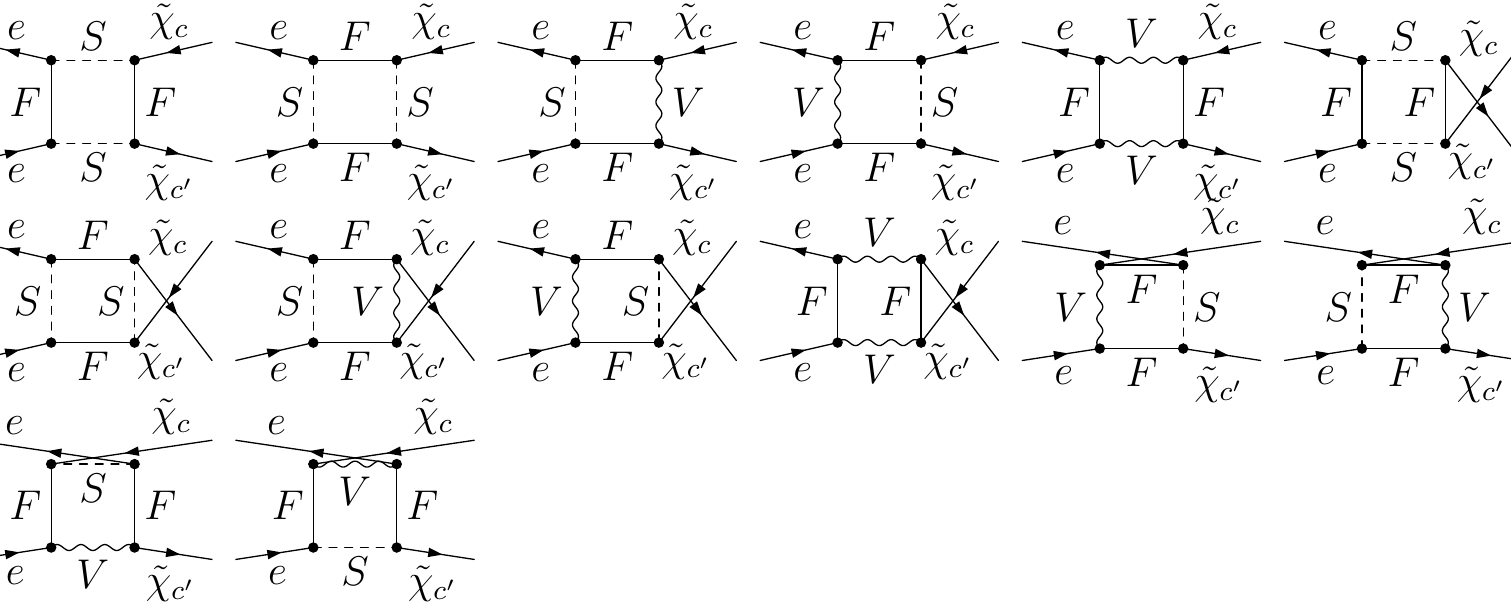}}
\framebox[14cm]{\includegraphics[width=0.67\textwidth]{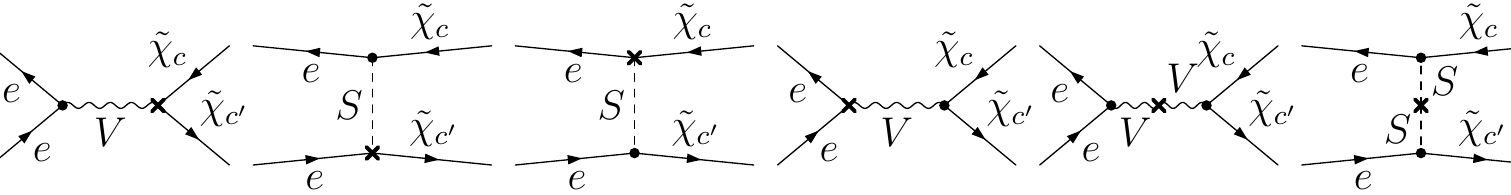}}
\caption{
  Generic tree, self-energy, vertex, box, and counterterm diagrams 
  for the process \eecc\ ($c,\cpri = 1,2$). 
  $F$ can be a SM fermion, chargino or neutralino; 
  $S$ can be a sfermion or a Higgs/Goldstone boson; 
  $V$ can be a $\ga$, $Z$ or $W^\pm$. 
  It should be noted that electron--Higgs couplings are neglected.  
}
\label{fig:CCiagrams}
\end{center}
\end{figure}
%%%%%%%%%%%%%%%%%%%%%%%%% F I G U R E %%%%%%%%%%%%%%%%%%%%%%%%%%%%%%%%%%%%%%%%%

%Fig2
%%%%%%%%%%%%%%%%%%%%%%%%% F I G U R E %%%%%%%%%%%%%%%%%%%%%%%%%%%%%%%%%%%%%%%%%
\begin{figure}
\begin{center}
\framebox[14cm]{\includegraphics[width=0.32\textwidth]{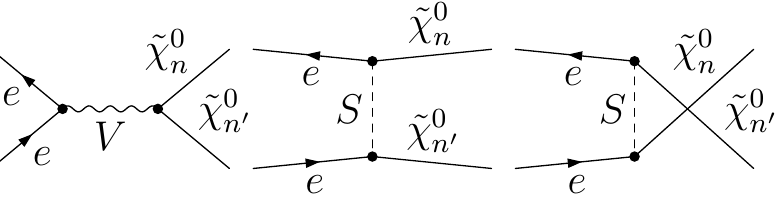}}
\framebox[14cm]{\includegraphics[width=0.67\textwidth]{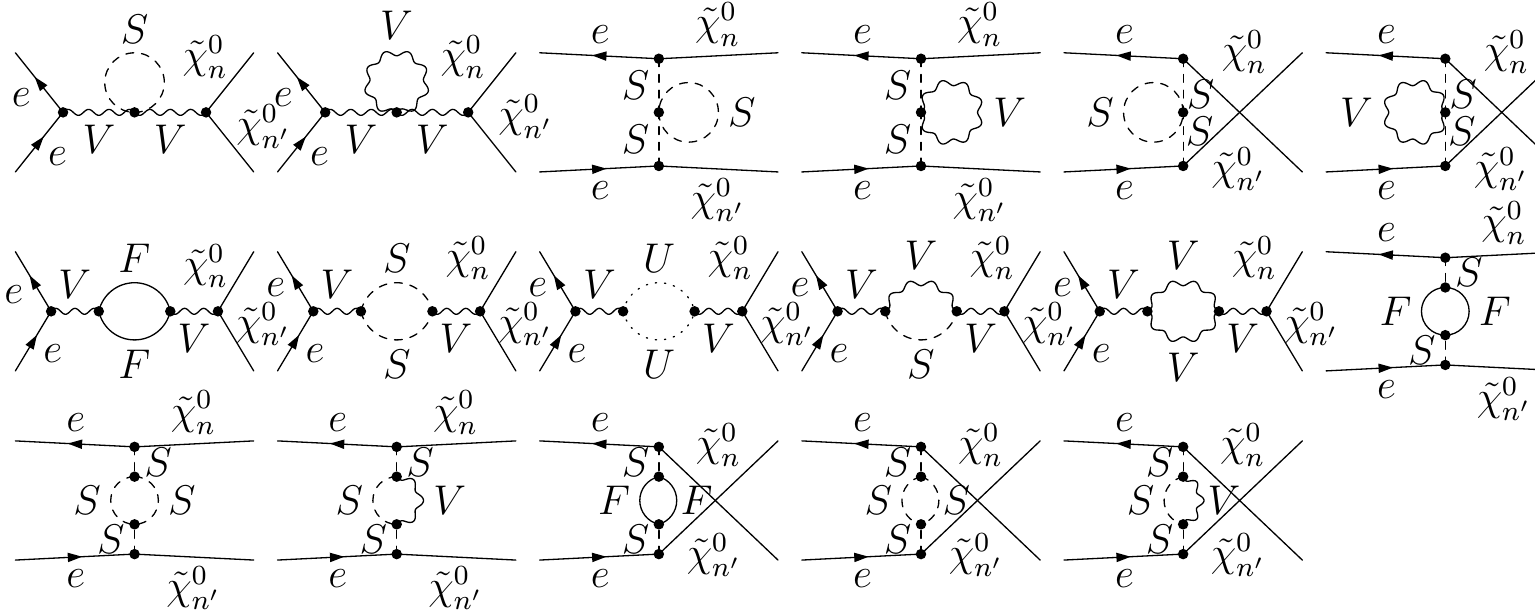}}
\framebox[14cm]{\includegraphics[width=0.67\textwidth]{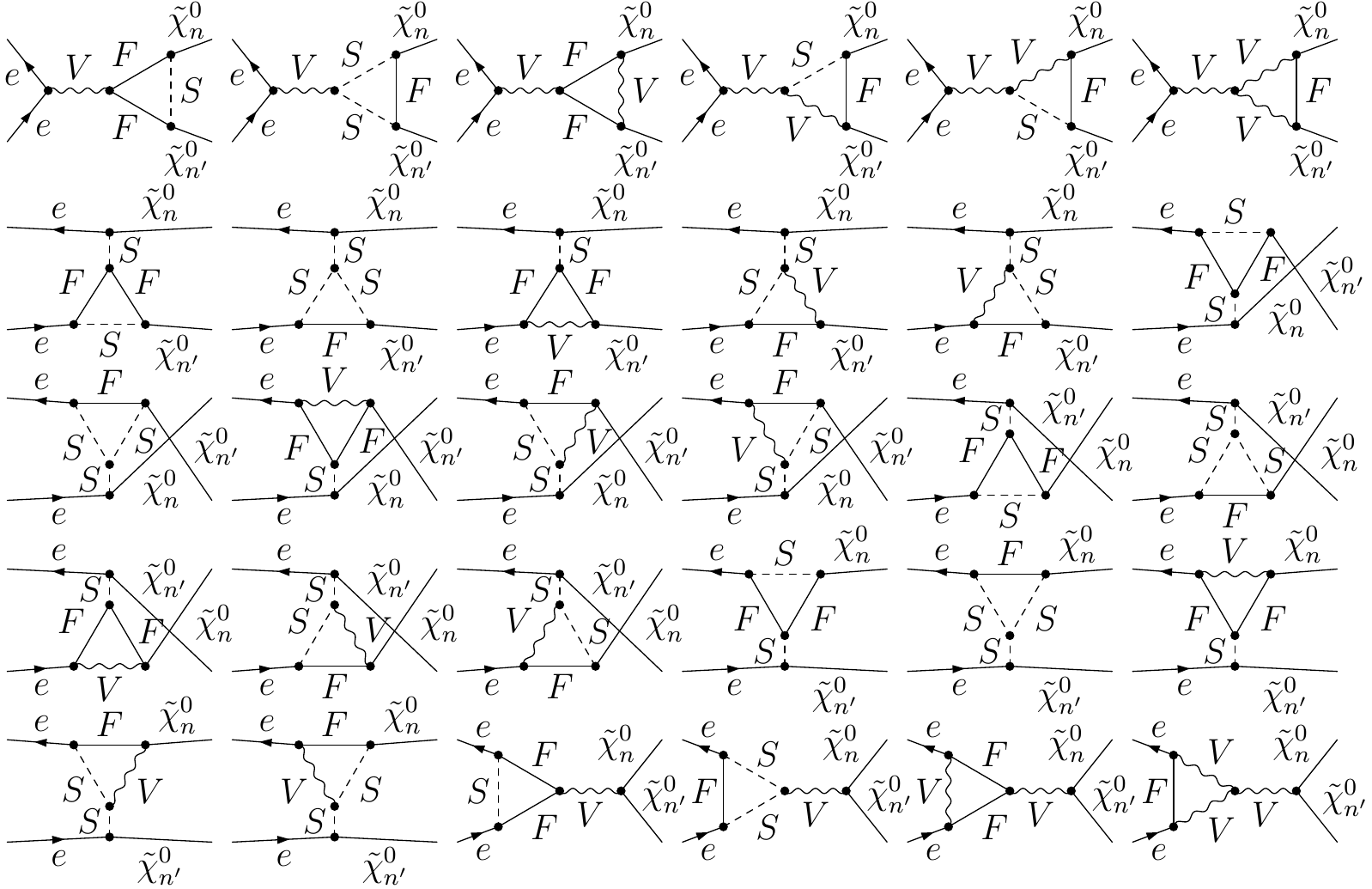}}
\framebox[14cm]{\includegraphics[width=0.67\textwidth]{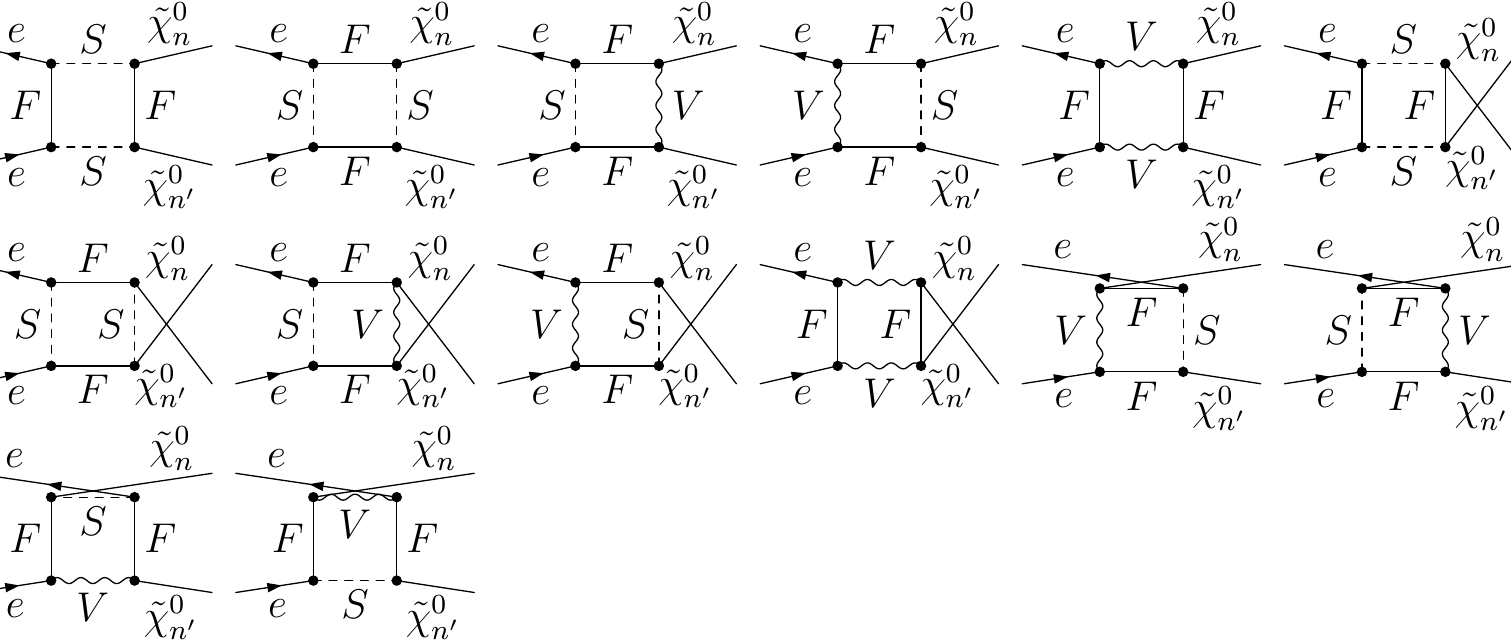}}
\framebox[14cm]{\includegraphics[width=0.67\textwidth]{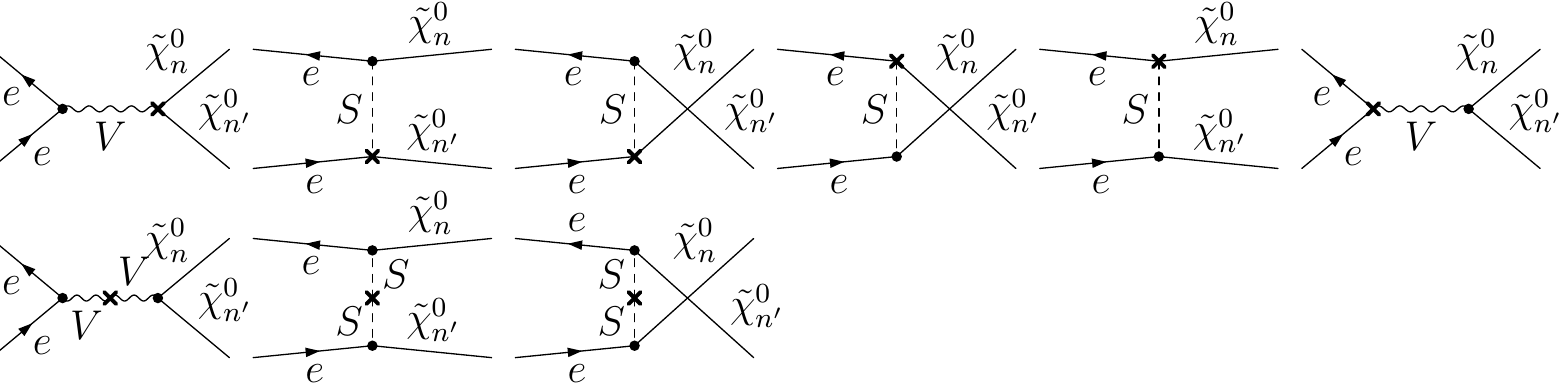}}
\caption{
  Same as \reffi{fig:CCiagrams}, but for the 
  process \eenn\ ($n,\npri = 1,2,3,4$). 
}
\label{fig:NNdiagrams}
\end{center}
\end{figure}
%%%%%%%%%%%%%%%%%%%%%%%%% F I G U R E %%%%%%%%%%%%%%%%%%%%%%%%%%%%%%%%%%%%%%%%%

Sample diagrams for the process \eecc\ are shown in \reffi{fig:CCiagrams} 
and for the process \eenn\ in \reffi{fig:NNdiagrams}.
Not shown are the diagrams for real (hard and soft) photon radiation. 
They are obtained from the corresponding tree-level diagrams by attaching a 
photon to the (incoming/outgoing) electron or chargino.
The internal particles in the generically depicted diagrams in 
\reffis{fig:CCiagrams} and \ref{fig:NNdiagrams} are labeled as follows: 
$F$ can be a SM fermion $f$, chargino $\cha{c}$ or neutralino 
$\neu{n}$; $S$ can be a sfermion $\Sf_s$ or a Higgs (Goldstone) boson 
$h^0, H^0, A^0, H^\pm$ ($G, G^\pm$); $U$ denotes the ghosts $u_V$;
$V$ can be a photon $\ga$ or a massive SM gauge boson, $Z$ or $W^\pm$. 
We have neglected all electron--Higgs couplings and terms proportional 
to the electron mass whenever this is safe, \ie except when the electron 
mass appears in negative powers or in loop integrals.
We have verified numerically that these contributions are indeed totally 
negligible.  For internally appearing Higgs bosons no higher-order
corrections to their masses or couplings are taken into account; 
these corrections would correspond to effects beyond one-loop order.

%%%%%%%%%%%%%%%%%%%%%%%%%%%%%%%%%%%%%%%%%%%%%%%%%%%%%%%%%%%%%%%%%%%%%%%%%%%%%%%

\subsection{Contributing diagrams for slepton production}
\label{sec:diagrams-slep}

\begin{figure}
\begin{center}
\framebox[14cm]{\includegraphics[width=0.24\textwidth]{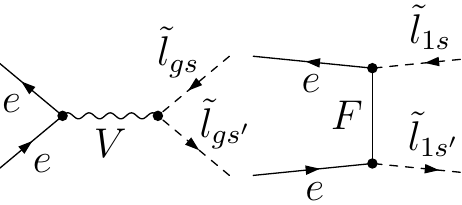}}
\framebox[14cm]{\includegraphics[width=0.75\textwidth]{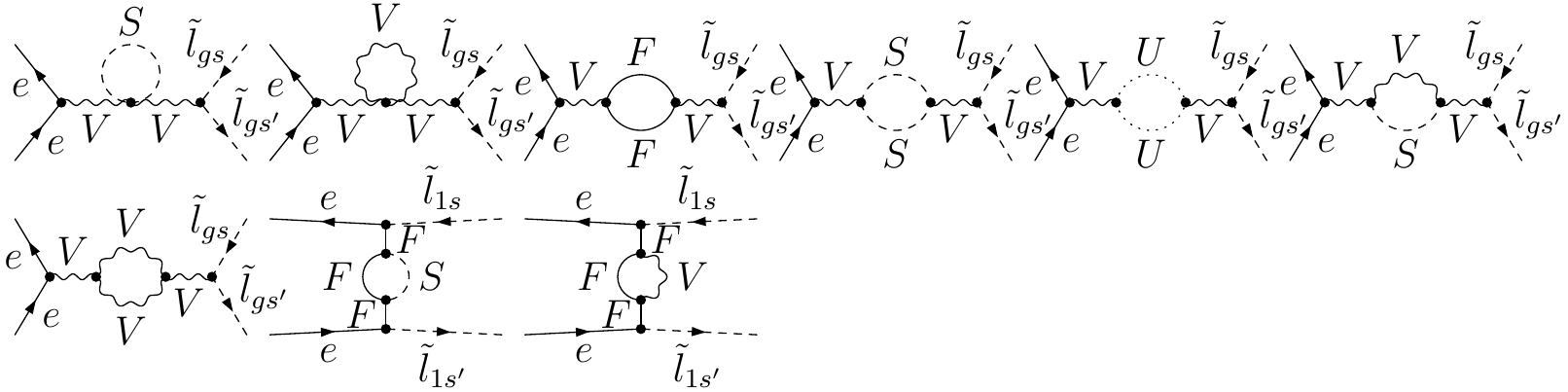}}
\framebox[14cm]{\includegraphics[width=0.75\textwidth]{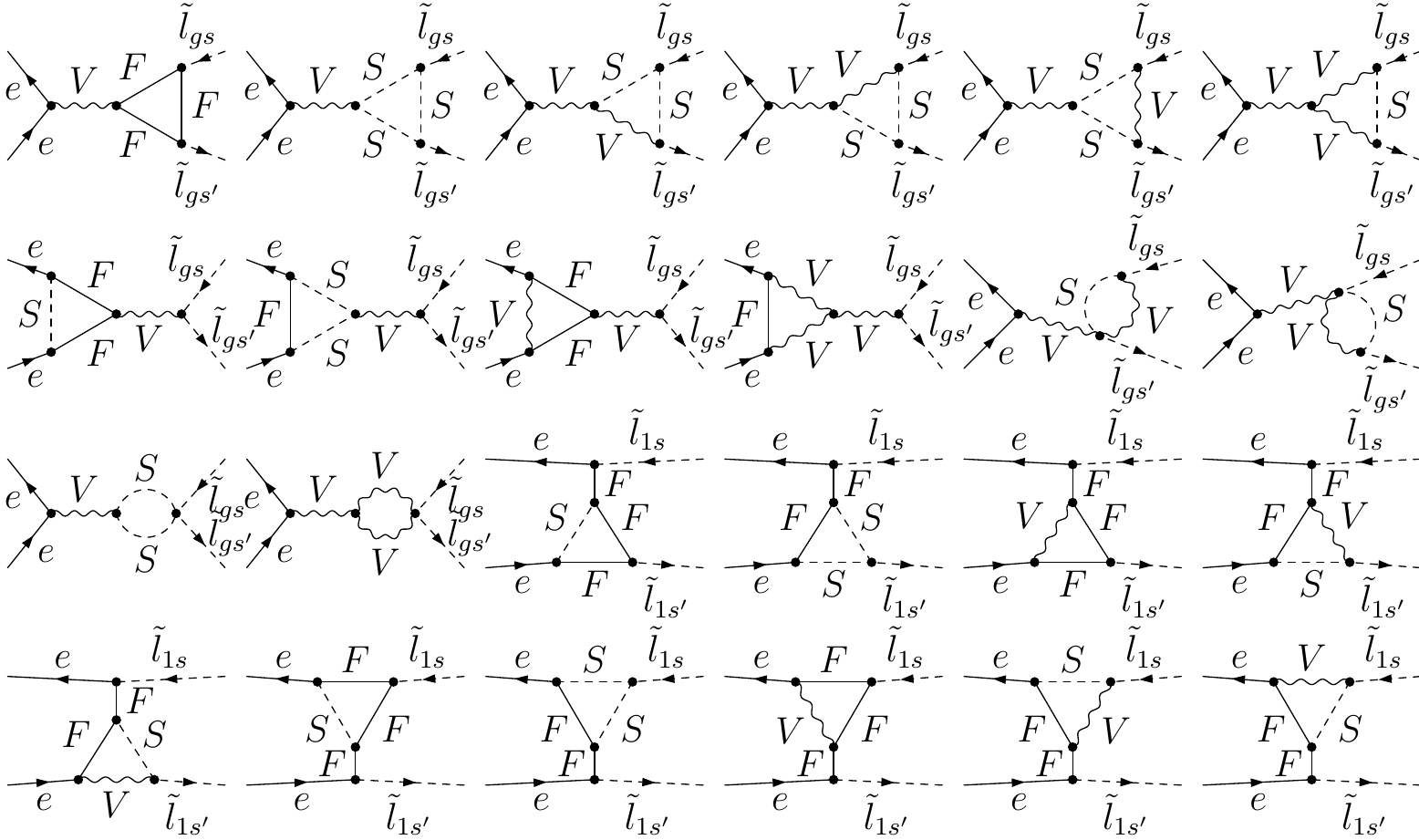}}
\framebox[14cm]{\includegraphics[width=0.75\textwidth]{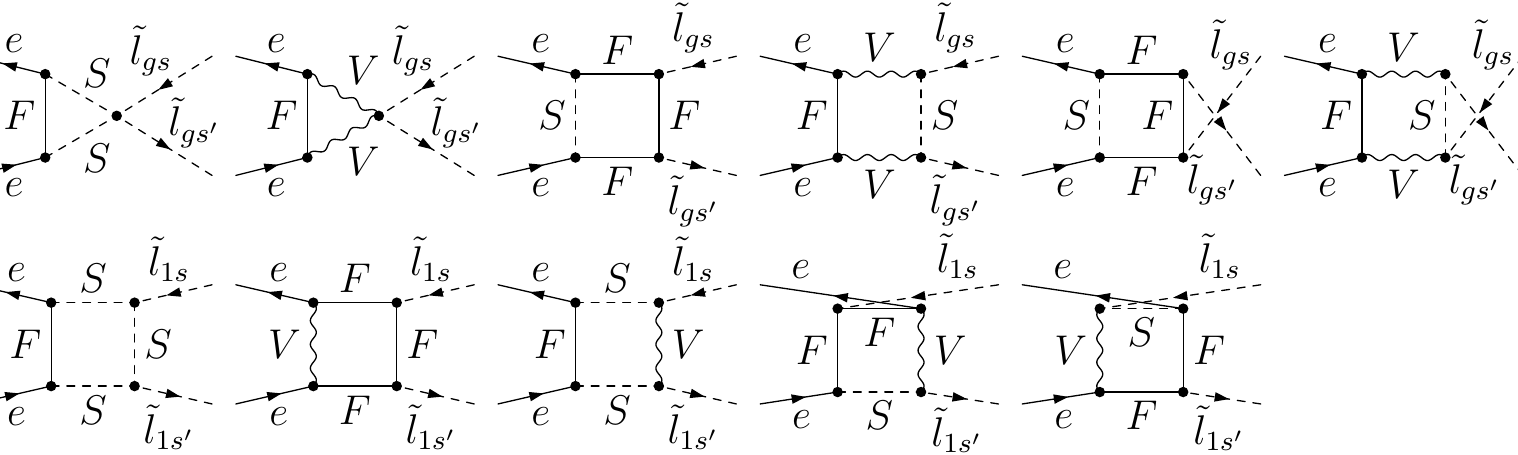}}
\framebox[14cm]{\includegraphics[width=0.75\textwidth]{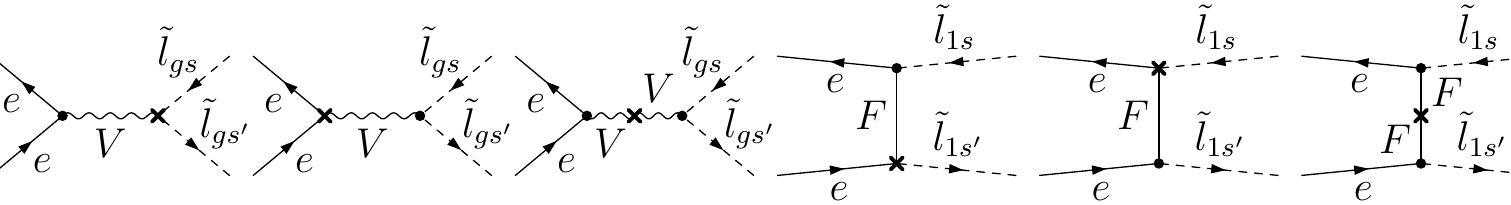}}
\caption{
  Generic tree, self-energy, vertex, box, and counterterm diagrams for the 
  process \eeSlSl\ ($\Sl_{gs} = \{\Se_{gs},\Sn_g\};\; g = 1,2,3;\; s,\spri = 1,2$). 
  The additional diagrams, which occur only in the case of first generation 
  slepton production, are denoted with $\Sl_{1s}$.
  $F$ can be a SM fermion, chargino or neutralino; 
  $S$ can be a sfermion or a Higgs/Goldstone boson; 
  $V$ can be a $\ga$, $Z$ or $W^\pm$. 
  It should be noted that electron--Higgs couplings are neglected. 
}
\label{fig:eeSlSl}
\end{center}
\end{figure}

Sample diagrams for the process \eeSeSe\ and \eeSnSn\ are shown in 
\reffi{fig:eeSlSl}. The diagrams not shown, the particle assignments and
the treatment of the Higgs sector are as in the previous subsection.

%Fig3
%%%%%%%%%%%%%%%%%%%%%%%%% F I G U R E %%%%%%%%%%%%%%%%%%%%%%%%%%%%%%%%%%%%%%%%%
%%%%%%%%%%%%%%%%%%%%%%%%% F I G U R E %%%%%%%%%%%%%%%%%%%%%%%%%%%%%%%%%%%%%%%%%

%%%%%%%%%%%%%%%%%%%%%%%%%%%%%%%%%%%%%%%%%%%%%%%%%%%%%%%%%%%%%%%%%%%%%%%%%%%%%
%%%%%%%%%%%%%%%%%%%%%%%%%%%%%%%%%%%%%%%%%%%%%%%%%%%%%%%%%%%%%%%%%%%%%%%%%%%%%

\section{Numerical analysis}
\label{sec:numeval}

Here we review the numerical analysis of chargino/neutralino
production at $e^+e^-$ colliders in the cMSSM as presented in
\citere{eeIno}. We also give a preview of the numerical analysis of
slepton production at $e^+e^-$ colliders that will be presented in
\citere{eeSlep}. 
In the figures below we show the cross sections at the tree level 
(``tree'') and at the full one-loop level (``full''), which is the cross 
section including \textit{all} one-loop corrections.
All results shown use the \Code{CCN[1]} renormalization scheme
(\ie OS conditions for the two charginos and the lightest neutralino).

The renormalization scale $\mu_R$ has been set to the center-of-mass energy, 
$\sqrt{s}$.  The SM parameters are chosen as follows; see also \cite{pdg}:
\begin{itemize}

\item Fermion masses (on-shell masses, if not indicated differently):
\begin{align}
m_e    &= 0.5109989461\mev\,, & m_{\nu_e}    &= 0\,, \notag \\
m_\mu  &= 105.6583745\mev\,,  & m_{\nu_{\mu}} &= 0\,, \notag \\
m_\tau &= 1776.86\mev\,,      & m_{\nu_{\tau}} &= 0\,, \notag \\
m_u &= 71.03\mev\,,          & m_d         &= 71.03\mev\,, \notag \\ 
m_c &= 1.27\gev\,,           & m_s         &= 96.0\mev\,, \notag \\
m_t &= 173.21\gev\,,         & m_b         &= 4.66\gev\,.
\end{align}
According to \citere{pdg}, $m_s$ is an estimate of a so-called 
"current quark mass" in the \MSbar\ scheme at the scale 
$\mu \approx 2\gev$.  $m_c \equiv m_c(m_c)$ is the "running" mass 
in the \MSbar\ scheme and $m_b$ is the $\Upsilon(1S)$ bottom quark mass. 
$m_u$ and $m_d$ are effective parameters, calculated through the 
hadronic contributions to
\begin{align}
\Delta\alpha_{\text{had}}^{(5)}(M_Z) &= 
      \frac{\alpha}{\pi}\sum_{f = u,c,d,s,b}
      Q_f^2 \Bigl(\ln\frac{M_Z^2}{m_f^2} - \frac 53\Bigr) \approx 0.02764\,.
\end{align}

\item Gauge-boson masses\index{gaugebosonmasses}:
\begin{align}
M_Z = 91.1876\gev\,, \qquad M_W = 80.385\gev\,.
\end{align}

\item Coupling constant\index{couplingconstants}:
\begin{align}
\alpha(0) = 1/137.035999139\,.
\end{align}
\end{itemize}

%%%%%%%%%%%%%%%%%%%%%%%%%%%%%%%%%%%%%%%%%%%%%%%%%%%%%%%%%%%%%%%%%%%%%%%%%%%%%%

\subsection{\texorpdfstring{The processes \boldmath{\eecc} and \boldmath{\eenn}}
                           {The process e+e- -> cha cha and e+e- -> neu neu}}
\label{sec:eeccnn}

%Tab1
%%%%%%%%%%%%%%%%%%%%% T A B L E %%%%%%%%%%%%%%%%%%%%%%%%%%%%%%%%%%%%%%%%%%%%%%
\begin{table}[t]
\caption{\label{tab:para-ccnn}
  MSSM default parameters for the numerical investigation of chargino
  and neutralino production; all parameters 
  (except of $\TB$) are in GeV.  The values for the trilinear sfermion 
  Higgs couplings, $A_{t,b,\tau}$ are chosen to be real and such that charge- 
  and/or color-breaking minima are avoided \cite{ccb}.  
  We have chosen common values for the three sfermion generations.
}
\centering
\begin{tabular}{lrrrrrrrrrrrr}
\toprule
Scen. & $\sqrt{s}$ & $\TB$ & $\mu$ & $\MHp$ & $M_{\tilde Q, \tilde U, \tilde D}$ & 
$M_{\tilde L, \tilde E}$ & $|\At|$ & $\Ab$ & $A_{\tau}$ & 
$|M_1|$ & $M_2$ & $M_3$ \\ 
\midrule
\Sce & 1000 & 10 & 450 & 500 & 1500 & 1500 & 2000 & $|\At|$ &
$M_{\tilde L}$ & $\mu$/4 & $\mu$/2 & 2000 \\
\bottomrule
\end{tabular}
\end{table}
%%%%%%%%%%%%%%%%%%%%% T A B L E %%%%%%%%%%%%%%%%%%%%%%%%%%%%%%%%%%%%%%%%%%%%%%

The SUSY parameters for the evaluation of these production cross sections 
are chosen according to the scenario \Sce, shown in \refta{tab:para-ccnn}.%
\footnote{
  It should be noted that changing $\mu$ also (by default) 
  changes $M_1$ and $M_2$ in our scenario \Sce.
}
This scenario is viable for the various cMSSM chargino/neutralino 
production modes, \ie not picking specific parameters for each cross 
section.  They are in particular in agreement with the chargino and 
neutralino searches of ATLAS~\cite{ATLAS-CN} and CMS~\cite{CMS-CN}. 

It should be noted that higher-order corrected Higgs boson masses do not 
enter our calculation.
However, we ensured that over larger parts of the parameter space the 
lightest Higgs boson mass is around $\sim 125 \pm 3\gev$ to indicate the
phenomenological validity of our scenarios. (The evaluation has been
done with the code \Code{\FH}~\cite{feynhiggs}.)
In the numerical evaluation in \citere{eeIno} the variations with $\sqrt{s}$,
$\mu$, $\MSL = \MSE$, and $\phiMe$, the phase of $M_1$ were analyzed.
In the following we show a few example results.

\begin{figure}[t]
\begin{center}
\begin{tabular}{c}
\includegraphics[width=0.48\textwidth,height=6cm]{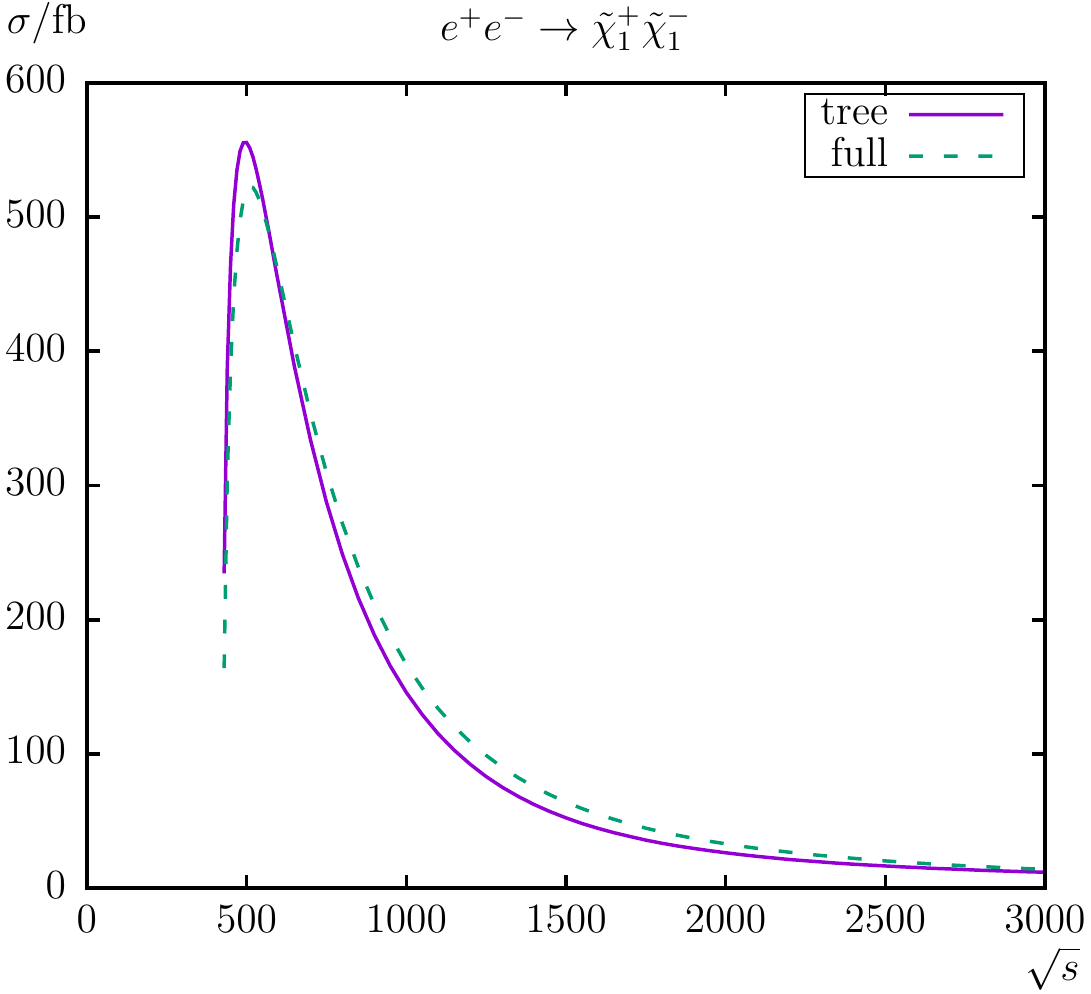}
\includegraphics[width=0.48\textwidth,height=6cm]{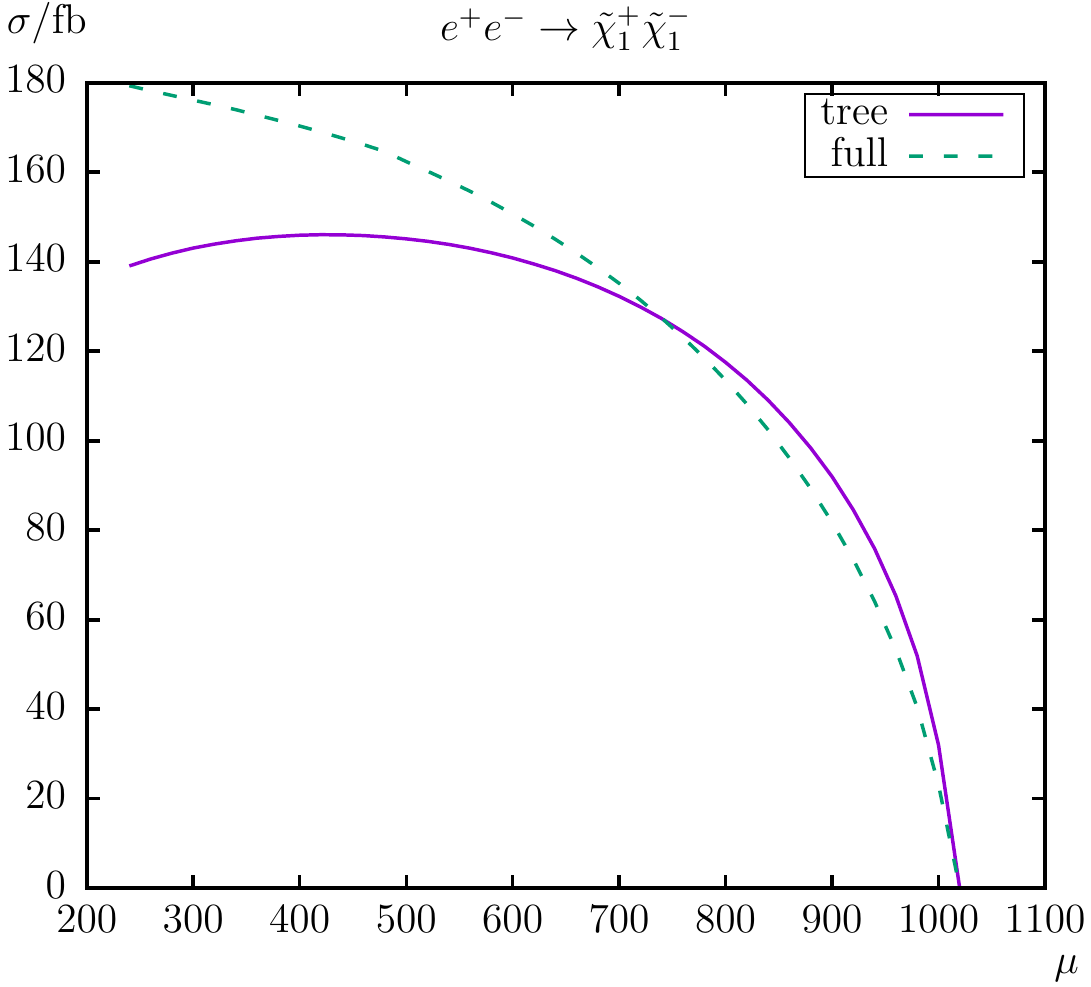}
\\[1em]
\includegraphics[width=0.48\textwidth,height=6cm]{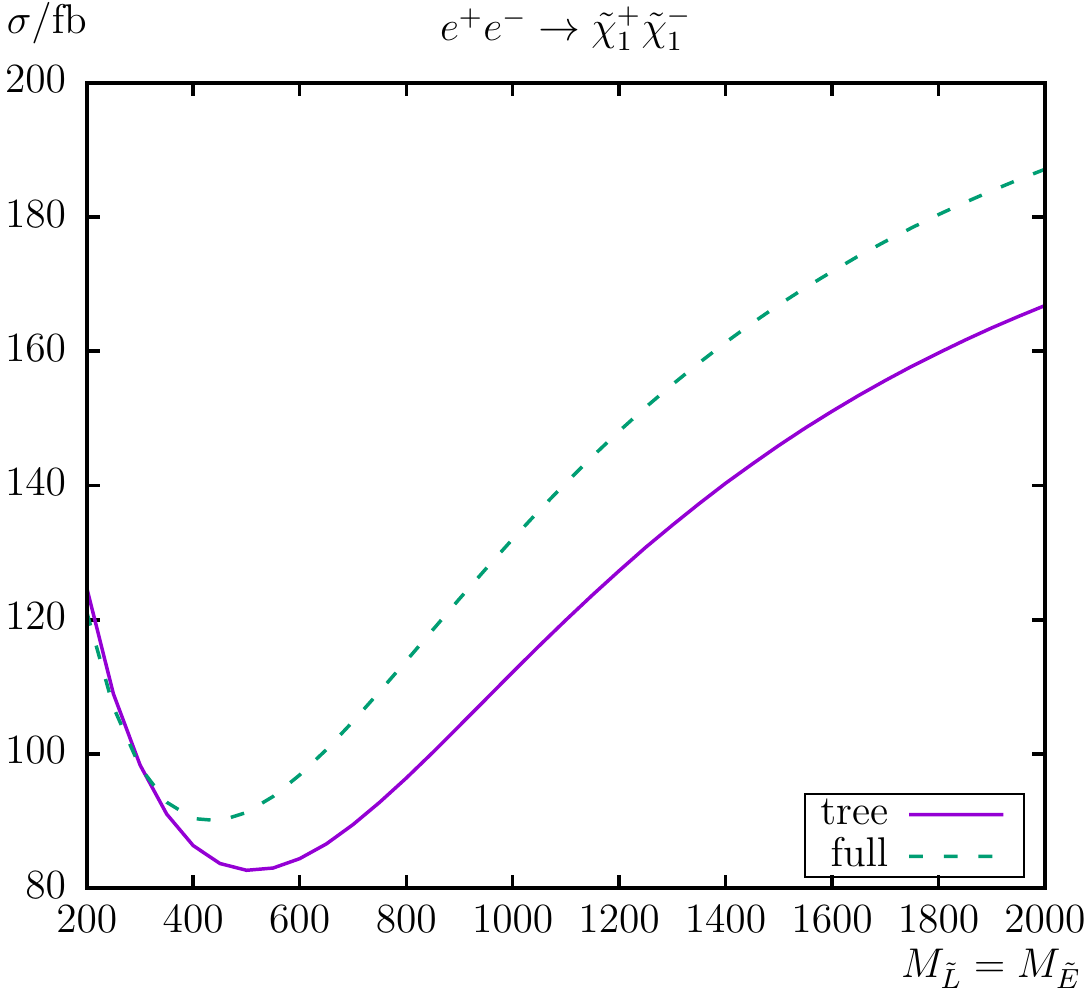}
\includegraphics[width=0.48\textwidth,height=6cm]{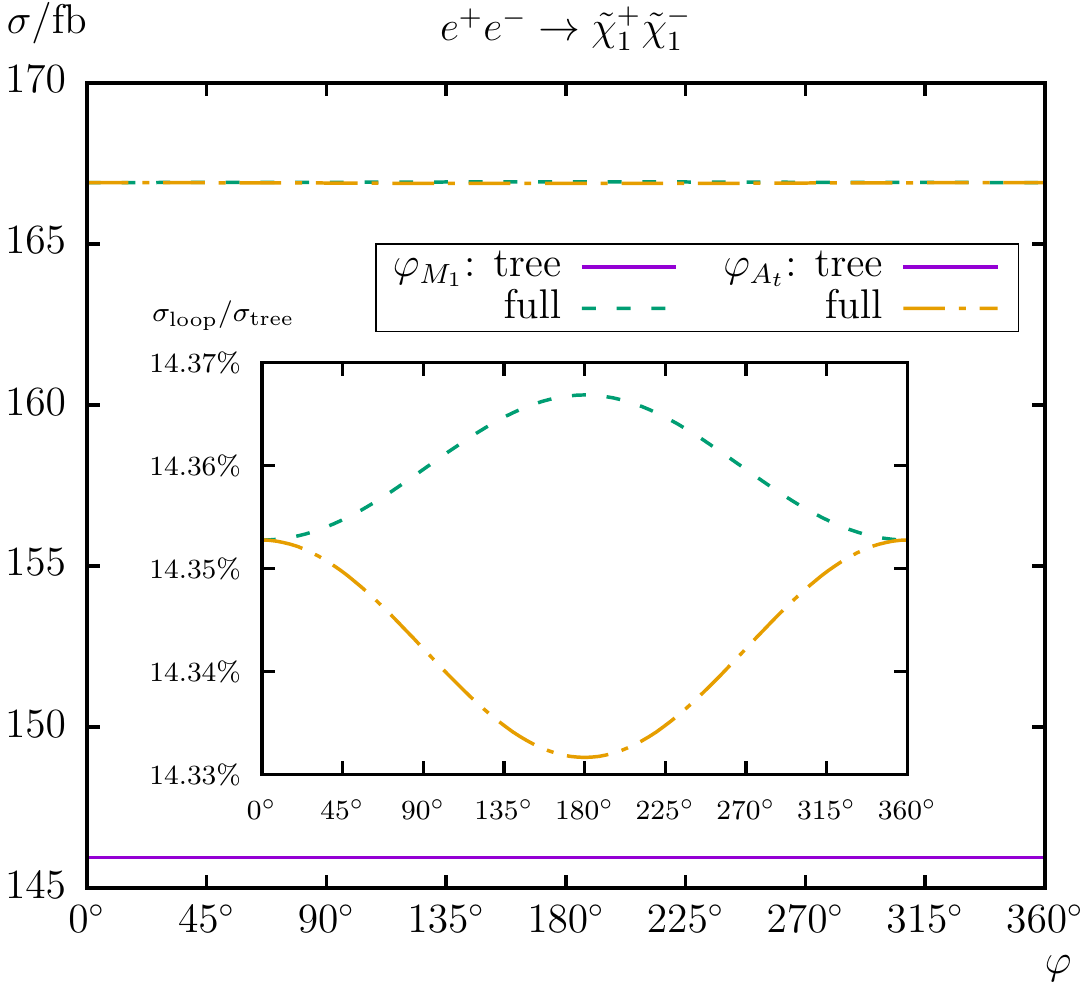}
\end{tabular}
\caption{\label{fig:eec1c1}
  $\sig(\eecece)$.
  Tree-level and full one-loop corrected cross sections are shown with 
  parameters chosen according to \Sce; see \refta{tab:para-ccnn}.
  The upper plots show the cross sections with $\sqrt{s}$ (left) and 
  $\mu$ (right) varied;  the lower plots show $\MSL = \MSE$ (left) and 
  $\phiMe$, $\phiAt$ (right) varied.
}
\end{center}
\end{figure}

\bigskip
The process $\eecece$ is shown in \reffi{fig:eec1c1}. 
It should be noted that for $s \to \infty$ decreasing cross sections 
$\propto 1/s$ are expected; see \citere{Bartl:1985fk}.
If not indicated otherwise, unpolarized electrons and positrons are 
assumed.

%Fig4
%%%%%%%%%%%%%%%%%%%%%%%%%% F I G U R E %%%%%%%%%%%%%%%%%%%%%%%%%%%%%%%%%%%%%%%
%%%%%%%%%%%%%%%%%%%%%%%%%% F I G U R E %%%%%%%%%%%%%%%%%%%%%%%%%%%%%%%%%%%%%%%

In the analysis of the production cross section as a function of $\sqrt{s}$ 
(upper left plot) we find the expected behavior: a strong rise close to the 
production threshold, followed by a decrease with increasing $\sqrt{s}$. 
We find a very small shift \wrt $\sqrt{s}$ around the production threshold
(not visible in the plot). 
Away from the production threshold, loop corrections of $\sim -8\,\%$ at 
$\sqrt{s} = 500\gev$ and $\sim +14\,\%$ at $\sqrt{s} = 1000\gev$ are found 
in scenario \Sce\ (see \refta{tab:para-ccnn}), with a ``tree crossing'' 
(\ie where the loop corrections become approximately zero and therefore 
cross the tree-level result) at $\sqrt{s} \approx 575\gev$.
The relative size of loop corrections increase with increasing $\sqrt{s}$ 
(and decreasing $\sig$) and reach $\sim +19\,\%$ at $\sqrt{s} = 3000\gev$.

With increasing $\mu$ in \Sce\ (upper right plot) we find a strong decrease 
of the production cross section, as can be expected from kinematics.
The relative loop corrections in \Sce\ reach $\sim +30\,\%$ 
at $\mu = 240\gev$ (at the border of the experimental limit), $\sim +14\,\%$ 
at $\mu = 450\gev$ (\ie \Sce) and $\sim -30\,\%$ at $\mu = 1000\gev$. 
In the latter case these large loop corrections are due to the (relative) 
smallness of the tree-level results, which goes to zero for $\mu = 1020\gev$
(\ie the chargino production threshold).

The cross section as a function of $\MSL$ ($= \MSE$) is shown in the lower 
left plot of \reffi{fig:eec1c1}.  This mass parameter controls the
$t$-channel exchange of first generation sleptons at tree-level.
First a small decrease down to $\sim 90$~fb can be observed for 
$\MSL \approx 400\gev$.  For larger $\MSL$ the cross section rises up 
to $\sim 190$~fb for $\MSL = 2\tev$.
In scenario \Sce\ we find a substantial increase of the cross sections from 
the loop corrections.  They reach the maximum of $\sim +18\,\%$ at 
$\MSL \approx 850\gev$ with a nearly constant offset of about $20$~fb
for higher values of $\MSL$.

Due to the absence of $\phiMe$ in the tree-level production cross section 
the effect of this complex phase is expected to be small.  Correspondingly
we find that the phase dependence $\phiMe$ of the cross section in our
scenario is tiny.  The loop corrections are found to be nearly independent 
of $\phiMe$ at the level below $\sim +0.1\,\%$ in \Sce.
We also show the variation with $\phiAt$, which enter via final state 
vertex corrections.  While the variation with $\phiAt$ is somewhat larger 
than with $\phiMe$, it remains tiny and unobservable.

The analyses for the production cross sections of $\sig(\eececz)$ and
$\sig(\eeczcz)$ can be found in \citere{eeIno}.
To summarize, for the chargino pair production a 
decreasing cross section $\propto 1/s$ for $s \to \infty$ was observed, 
see \citere{Bartl:1985fk}. 
The full one-loop corrections are very roughly 10-20\,\% of the
tree-level results, but depend strongly on the size of $\mu$, where
larger values result even in negative loop corrections.
The cross sections are largest for $\eecece$ and $\eeczcz$ and roughly
smaller by one order of magnitude for $\eececz$. 
This is because there is no $\gamma\, \chapm{1} \champ{2}$ coupling at 
tree level in the MSSM.

\begin{figure}[t]
\begin{center}
\begin{tabular}{c}
\includegraphics[width=0.48\textwidth,height=6cm]{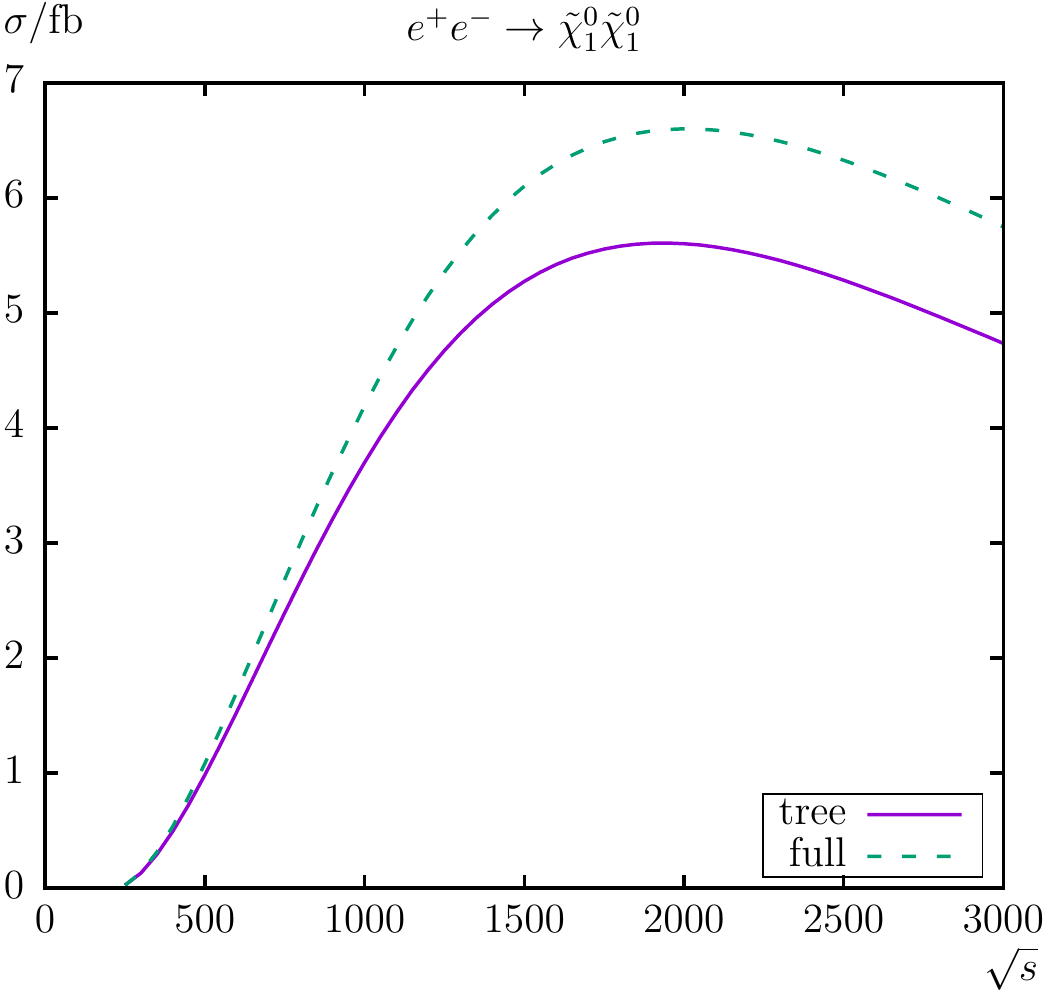}
\includegraphics[width=0.48\textwidth,height=6cm]{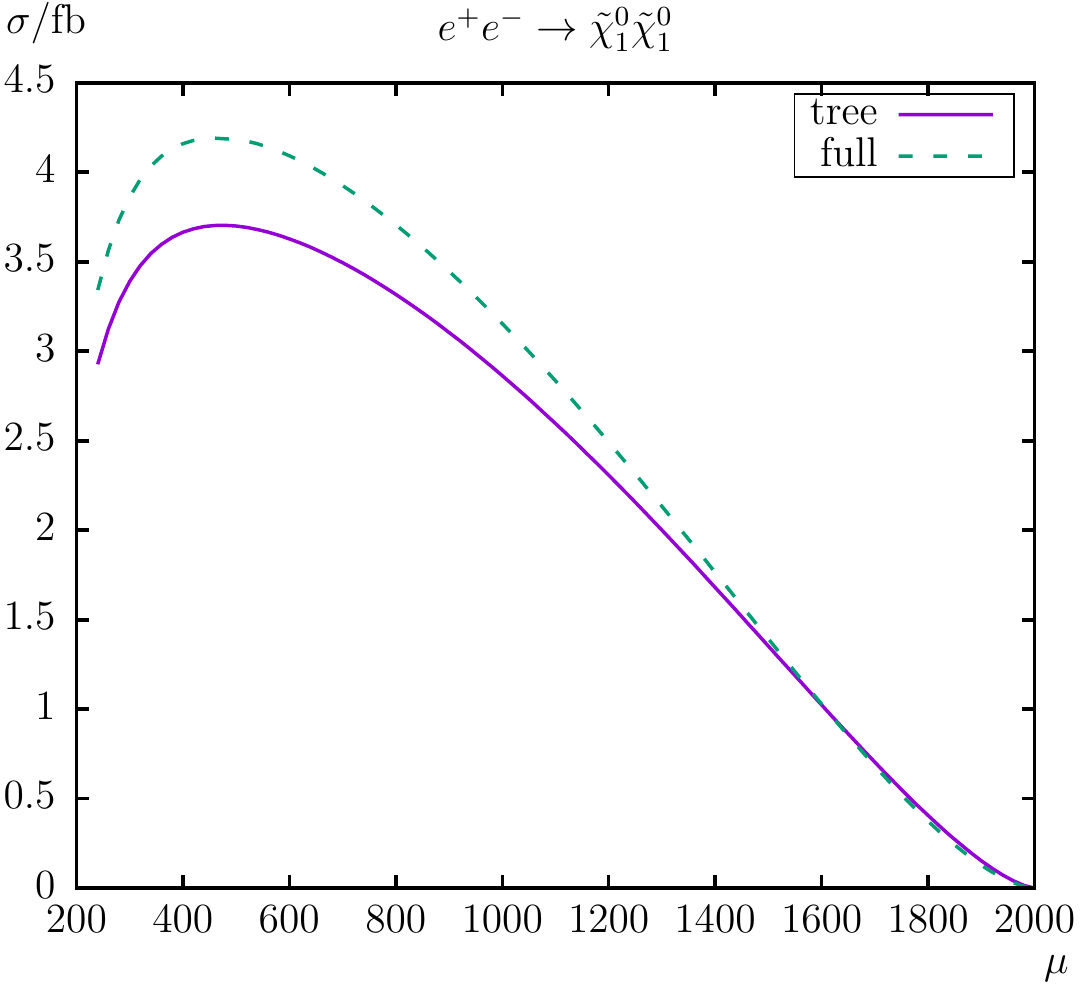}
\\[1em]
\includegraphics[width=0.48\textwidth,height=6cm]{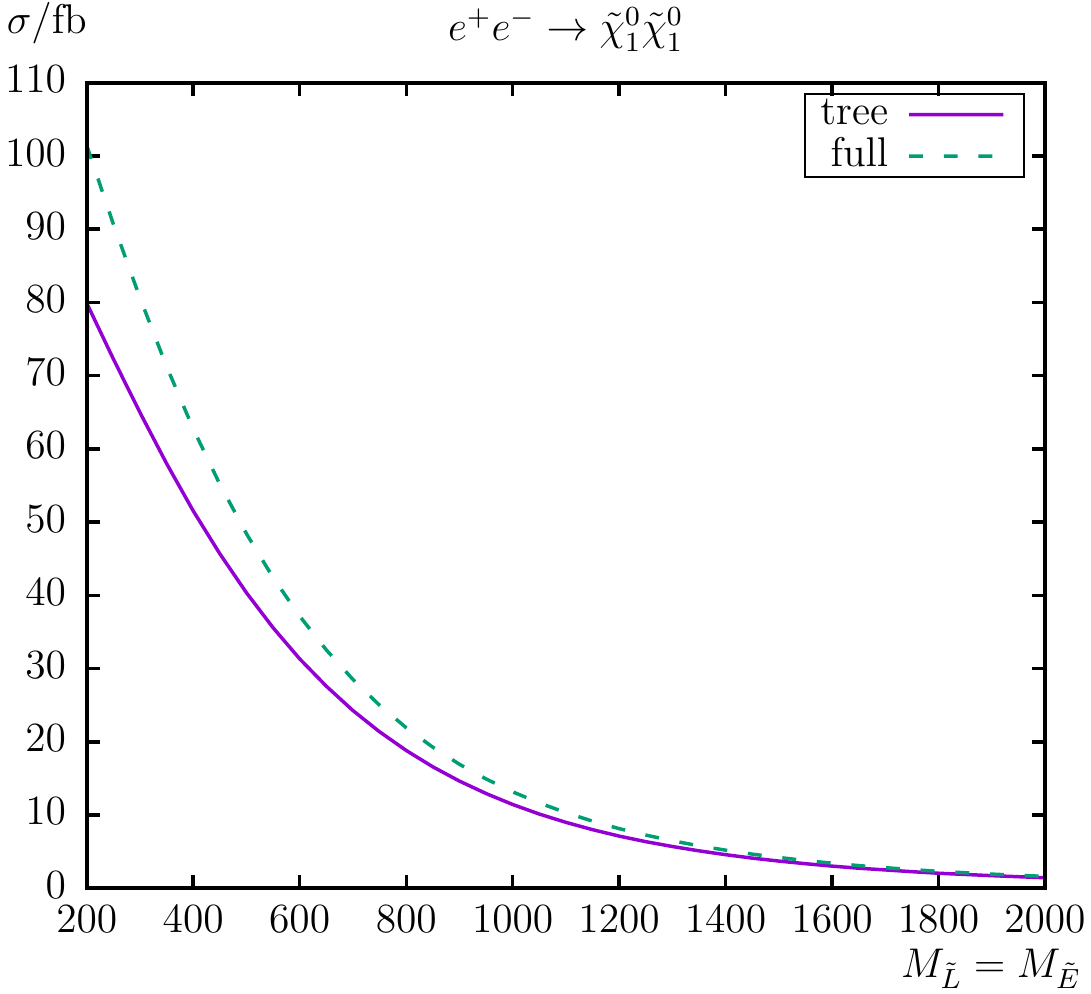}
\includegraphics[width=0.48\textwidth,height=6cm]{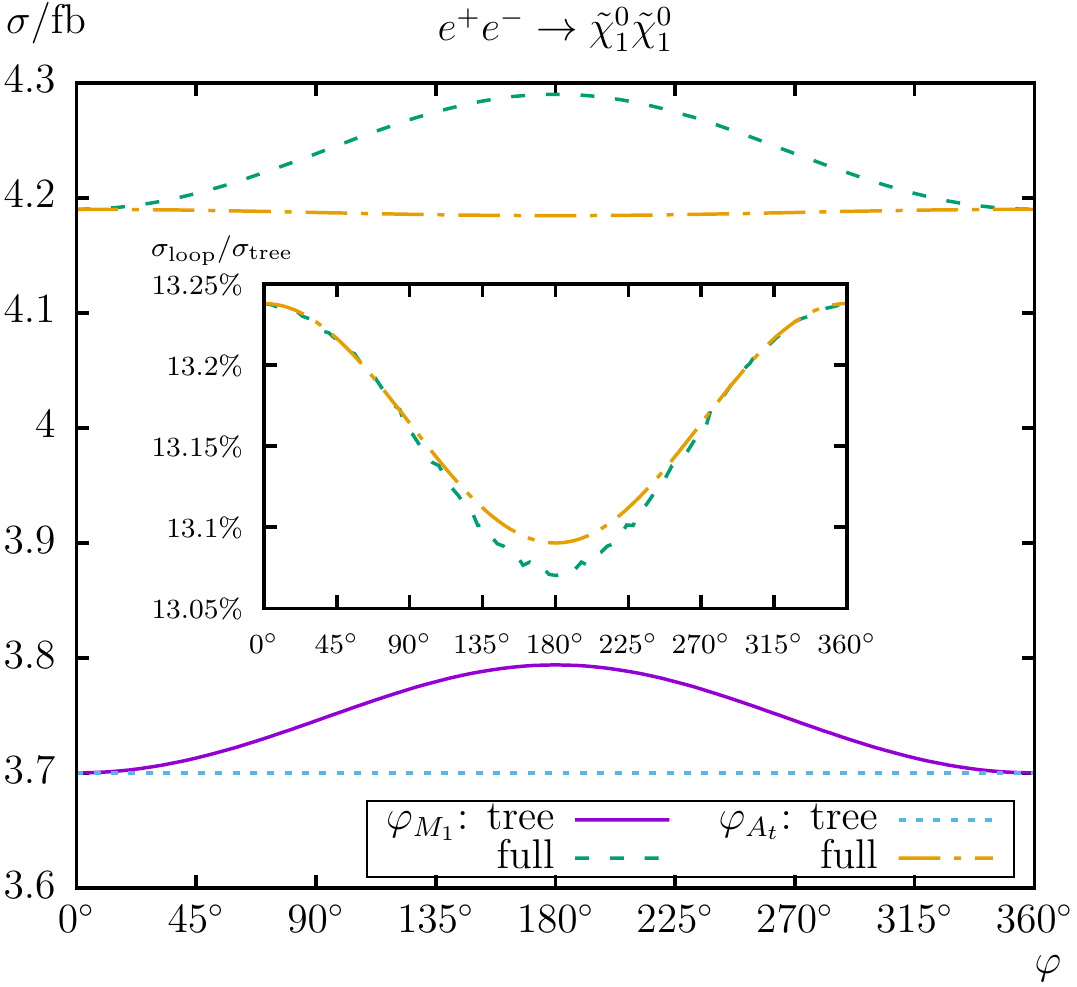}
\end{tabular}
\caption{\label{fig:een1n1}
  $\sig(\eenene)$.
  Tree-level and full one-loop corrected cross sections are shown with 
  parameters chosen according to \Sce; see \refta{tab:para-ccnn}.
  The upper plots show the cross sections with $\sqrt{s}$ (left) and 
  $\mu$ (right) varied;  the lower plots show $\MSL = \MSE$ (left) and 
  $\phiMe$, $\phiAt$ (right) varied.
}
\end{center}
\end{figure}

\bigskip
We now turn to the neutralino production cross sections. First, 
the process $\eenene$ is shown in \reffi{fig:een1n1}.
Away from the production threshold, loop corrections of $\sim +13\,\%$ at 
$\sqrt{s} = 1000\gev$ are found in scenario \Sce\ (see \refta{tab:para-ccnn}), 
with a maximum of nearly 7~fb at $\sqrt{s} \approx 2000\gev$. 
The relative size of the loop corrections increase with increasing 
$\sqrt{s}$ and reach $\sim +22\,\%$ at $\sqrt{s} = 3000\gev$.

With increasing $\mu$ in \Sce\ (upper right plot) we find a strong decrease 
of the production cross section, as can be expected from kinematics, 
discussed above.  The relative loop corrections reach $\sim +14\,\%$ at 
$\mu = 240\gev$ (at the border of the experimental exclusion bounds) and 
$\sim +13\,\%$ at $\mu = 450\gev$ (\ie \Sce).
The tree crossing takes place at $\mu \approx 1600\gev$.  For higher
$\mu$ values the loop corrections are negative, where the relative size
becomes large due to the (relative) smallness of the tree-level results,
which goes to zero for $\mu \approx 2000\gev$.

%Fig5
%%%%%%%%%%%%%%%%%%%%%%%%%% F I G U R E %%%%%%%%%%%%%%%%%%%%%%%%%%%%%%%%%%%%%%%
%%%%%%%%%%%%%%%%%%%%%%%%%% F I G U R E %%%%%%%%%%%%%%%%%%%%%%%%%%%%%%%%%%%%%%%

The cross sections are decreasing with increasing $\MSL$, \ie the (negative)
interference of the $t$-channel exchange decreases the cross sections, 
and the full one-loop result has its maximum of $\sim 100$~fb at 
$\MSL = 200\gev$. 
Analogously the relative corrections are decreasing from $\sim +27\,\%$ at 
$\MSL = 200\gev$ to $\sim +12\,\%$ at $\MSL = 2000\gev$.  For the other 
parameter variations one can conclude that a cross section larger by nearly 
one order of magnitude can be possible for very low $\MSL$ 
(which are not yet excluded experimentally).

Now we turn to the complex phase dependence. As for the chargino production, 
$\phiAt$ enters only via final state vertex corrections.  On the other hand, 
$\phiMe$ enters already at tree-level, and correspondingly larger effects 
are expected.
We find that the phase dependence $\phiMe$ of the cross section in \Sce\ 
is small (lower right plot), possibly not completely negligible, 
amounting up to $\sim 2.3\,\%$ for the full corrections.  
The loop corrections at the level of $\sim +13\,\%$ are found to be nearly 
independent of $\phiMe$, with a relative variation of $\sigloop/\sigtree$ 
at the level of $\sim +0.2\,\%$, 
(see the inlay in the lower right plot of \reffi{fig:een1n1}).
The loop effects of $\phiAt$ are found at the same level as the ones of 
$\phiMe$, \ie rather negligible.

%Fig6
%%%%%%%%%%%%%%%%%%%%%%%%%% F I G U R E %%%%%%%%%%%%%%%%%%%%%%%%%%%%%%%%%%%%%%%
\begin{figure}
\begin{center}
\begin{tabular}{c}
\includegraphics[width=0.48\textwidth,height=6cm]{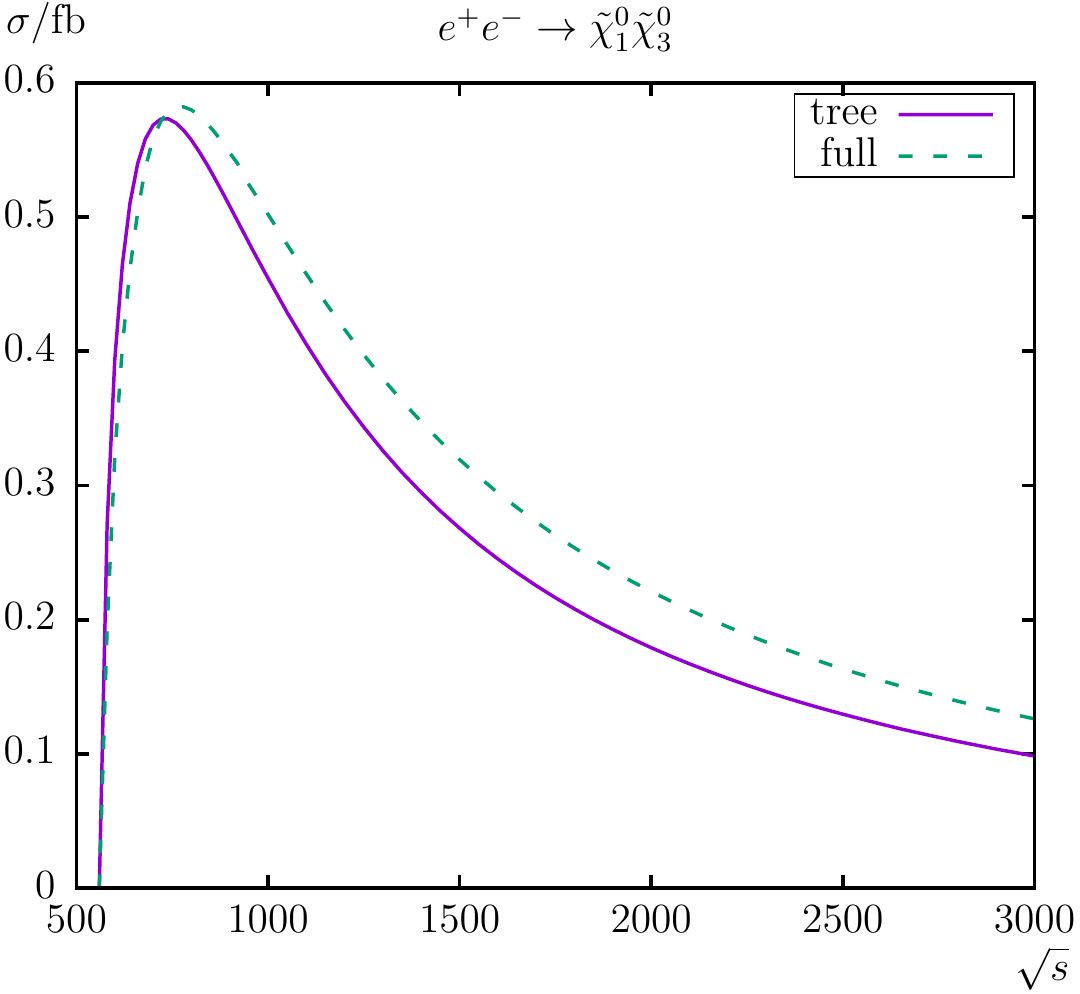}
\includegraphics[width=0.48\textwidth,height=6cm]{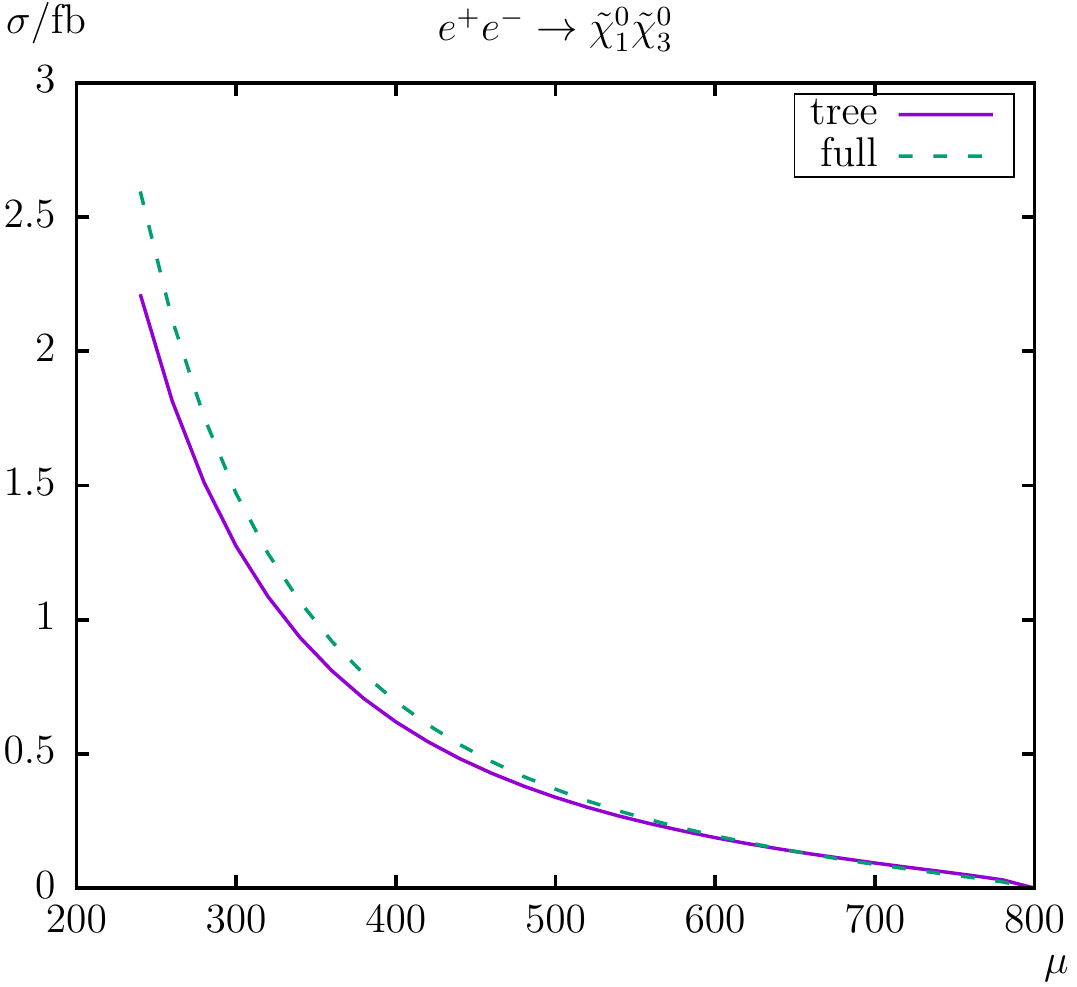}
\\[1em]
\includegraphics[width=0.48\textwidth,height=6cm]{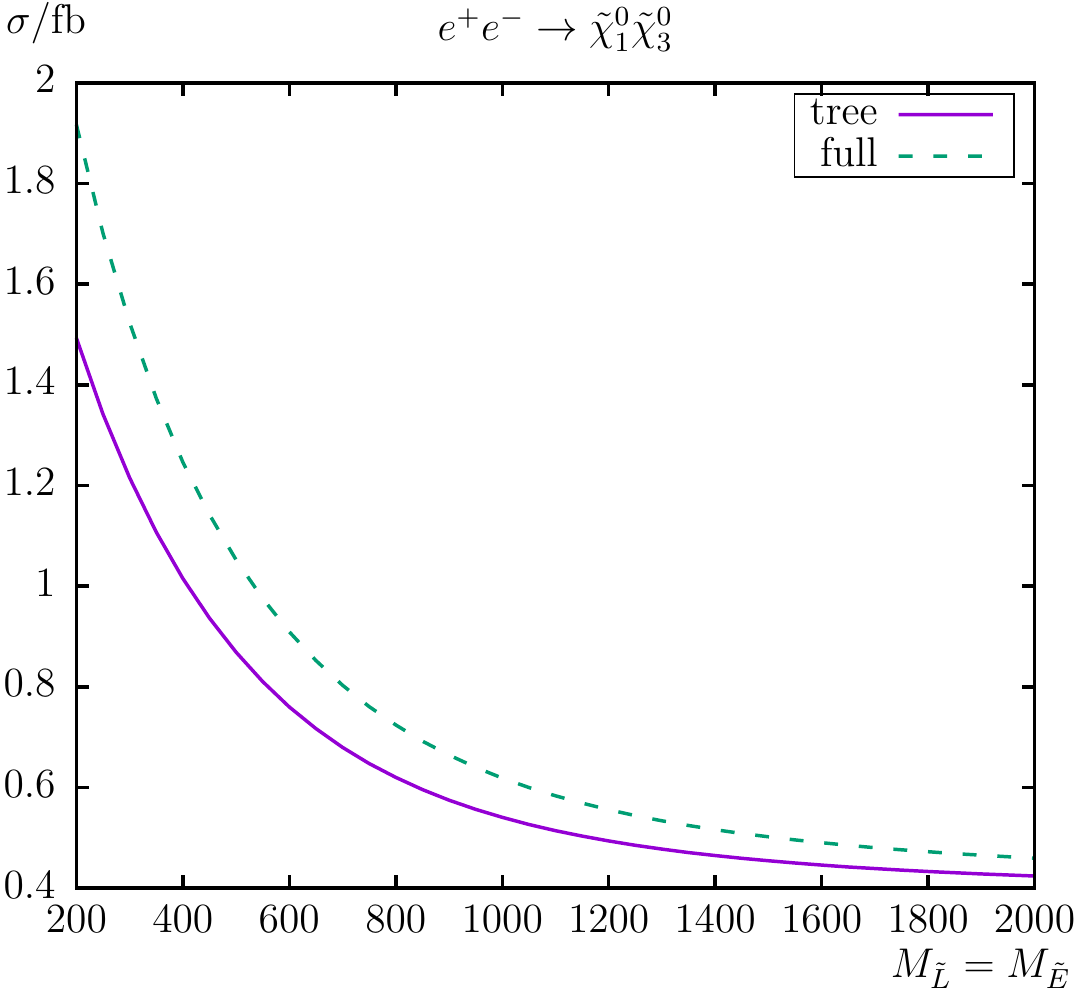}
\includegraphics[width=0.48\textwidth,height=6cm]{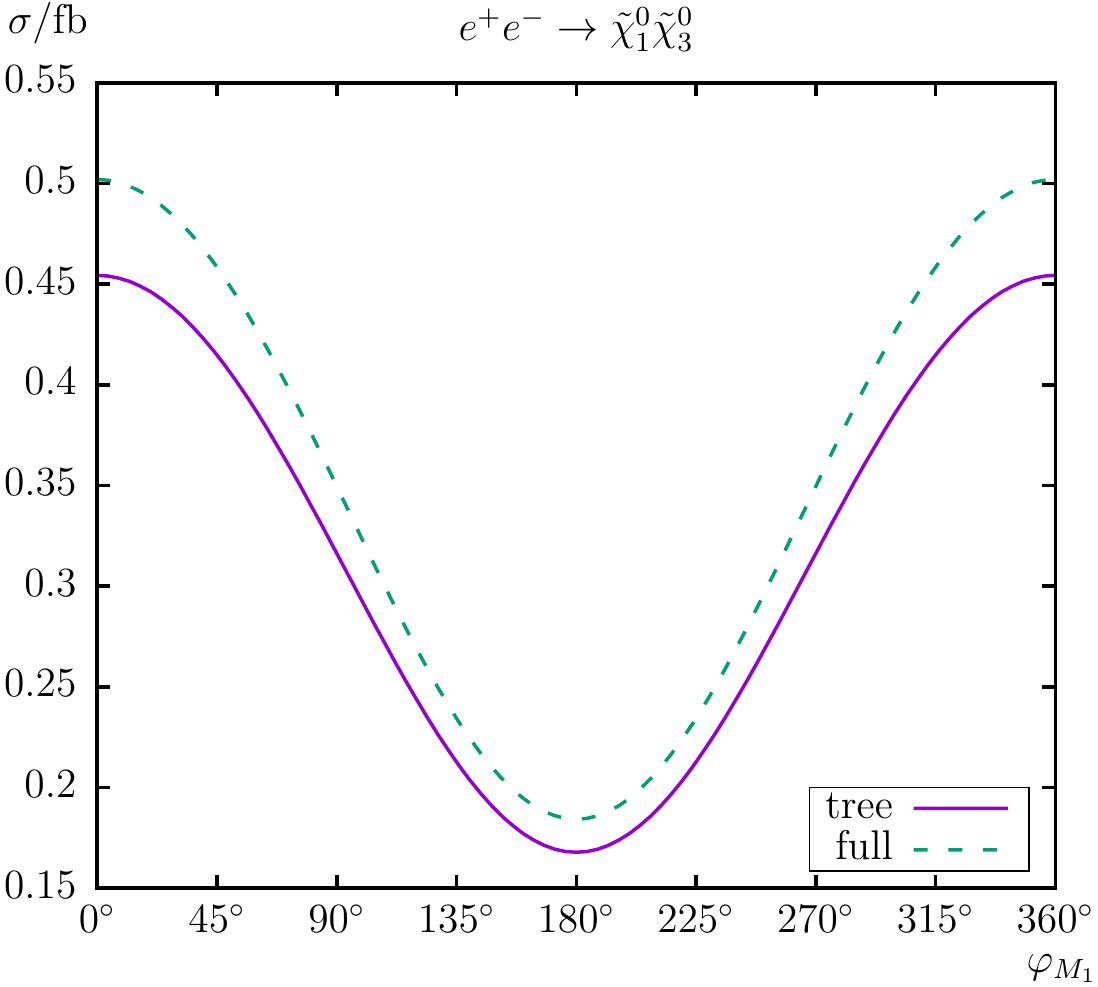}
\\[1em]
\includegraphics[width=0.48\textwidth,height=6cm]{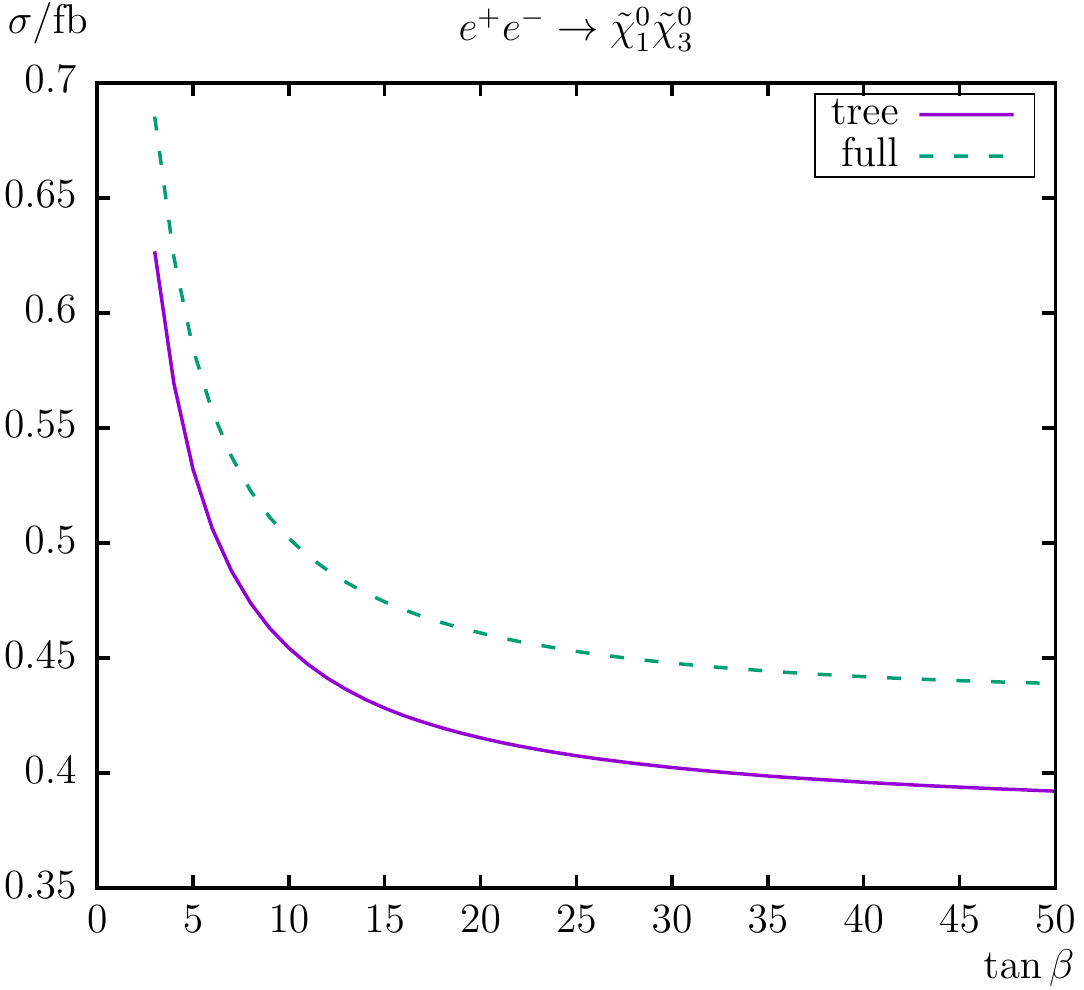}
\end{tabular}
\caption{\label{fig:een1n3}
  $\sig(\eenend)$.
  Tree-level and full one-loop corrected cross sections are shown with 
  parameters chosen according to \Sce; see \refta{tab:para-ccnn}.
  The upper plots show the cross sections with $\sqrt{s}$ (left) and 
  $\mu$ (right) varied;  the middle plots show $\MSL = \MSE$ (left) and 
  $\phiMe$ (right) varied; the lower plot shows the variation with $\TB$.
}
\end{center}
\end{figure}
%%%%%%%%%%%%%%%%%%%%%%%%%% F I G U R E %%%%%%%%%%%%%%%%%%%%%%%%%%%%%%%%%%%%%%%

%\bigskip
Second, the process $\eenend$ shown in \reffi{fig:een1n3},
which is found to be rather small of \order{1\, \fb}.
As a function of $\sqrt{s}$ (upper row, left plot) we find a small 
shift \wrt $\sqrt{s}$ directly at the production threshold, as well 
as a shift of $\sim +50\gev$ of the maximum cross section position.
The loop corrections range from $\sim +11\,\%$ at $\sqrt{s} = 1000\gev$ 
(\ie \Sce) to $\sim +28\,\%$ at $\sqrt{s} = 3000\gev$.

The dependence on $\mu$ (upper right plot) is rather small.  The relative 
corrections are $\sim +17\,\%$ at $\mu = 240\gev$, $\sim +11\,\%$ at 
$\mu = 450\gev$ (\ie \Sce), and have a tree crossing at
$\mu \approx 650\gev$. 
For larger $\mu$ the cross section goes to zero due to kinematics.

The cross section decreases with $\MSL$ (middle left plot), again
due to the negative interference of the $t$-channel
contribution. The full correction has a maximum of $\sim 2$~fb for 
$\MSL = 200\gev$, going down to $\sim 0.5$~fb at $\MSL = 2000\gev$.
Analogously the relative corrections are decreasing from $\sim +28\,\%$ 
at $\MSL = 200\gev$ to $\sim +8\,\%$ at $\MSL = 2000\gev$.

The phase dependence $\phiMe$ of the cross section in \Sce\ is shown in 
the middle right plot of \reffi{fig:een1n3}.  It is very pronounced
and can vary $\sigfull(\eenend)$ by 60\,\%.
The (relative) loop corrections are at the level of $\sim 10\,\%$ \wrt 
the tree cross section. 

Here we also show the variation with $\TB$ in the lower plot of 
\reffi{fig:een1n3}. 
The loop corrected cross section decreases from $\sim 0.7$~fb at small
$\TB$ to $\sim 0.45$~fb at $\TB = 50$.  The relative corrections for the 
$\TB$ dependence are increasing from $\sim +9\,\%$ at $\TB = 3$ to 
$\sim +12\,\%$ at $\TB = 50$.

The numerical analysis of the other neutralino production cross sections
can be found in \citere{eeIno}. 
To summarize, for the neutralino pair production the leading order corrections 
can reach a level of \order{10\, \fb}, depending on the SUSY parameters,
but is very small for the production of two equal higgsino dominated
neutralinos at the \order{10\, \ab} level.  This renders these 
processes difficult to observe at an $e^+e^-$ collider.%
\footnote{
  The limit of $10$~ab corresponds to ten events at an integrated 
  luminosity of $\cL = 1\, \iab$, which constitutes a guideline 
  for the observability of a process at a linear collider.
}
Having both beams polarized could turn out to be crucial to yield a
detectable production cross section in this case; 
see \citere{pol-report} for related analyses.

The full one-loop corrections are very roughly 10-20\,\% of the tree-level 
results, but vary strongly on the size of $\mu$ and $\MSL$.
Depending on the size of in particular these two parameters the loop 
corrections can be either positive or negative.
This shows that the loop corrections, while being large, have to be
included point-by-point in any precision analysis.
The dependence on $\phiMe$ was found at the level of $\sim 15\,\%$, but
can go up to $\sim 40\,\%$ for the extreme cases.  
The relative loop corrections varied by up to $5\,\%$ with $\phiMe$.
Consequently, the complex phase dependence must be taken into account 
as well. 

%%%%%%%%%%%%%%%%%%%%%%%%%%%%%%%%%%%%%%%%%%%%%%%%%%%%%%%%%%%%%%%%%%%%%%%%%%%%%%%

\subsection{\texorpdfstring{The processes \boldmath{\eeSeSe} and \boldmath{\eeSnSn}}
                           {The process e+e- -> Slepton Slepton and e+e- -> Sneutrino Sneutrino}}
\label{sec:eeslepslep}

The SUSY parameters for the numerical analysis here (\ie\ in
\citere{eeSlep}) are chosen according to the scenario \Scz, shown 
in \refta{tab:para-slep}.  This scenario is viable for the various cMSSM 
slepton production modes, again not picking specific parameters for 
each cross section.  They are in particular in agreement with 
the relevant SUSY searches of ATLAS and CMS:
Our electroweak spectrum is not covered by the latest ATLAS/CMS exclusion 
bounds, where two limits have to be distinguished. The limits not taking into
account a possible intermediate slepton exclude a lightest neutralino only 
well below $300\gev$~\cite{ATLAS1,CMS1}, whereas in \Scz\ we have 
$\mneu1 \approx 323\gev$.  Limits with intermediary sleptons often assume 
a chargino decay to lepton and sneutrino, while in our scenario 
$\mcha1 < \msneu$. Furthermore, the exclusion bounds given in the 
$\mneu1$-$\mneu2$ mass plane (with $\mneu2 \approx \mcha1$ assumed) above 
$\mneu2 \sim 300 \gev$ do not cover a compressed spectrum~\cite{ATLAS1,CMS2} 
for $\neu1$, $\neu2$, and $\cha1$.  In particular our scenario \Scz\ assumed 
masses of $\mneu1 \approx 323\gev$ and $\mneu2 \approx 354\gev$, which are 
not excluded.

%Tab2
%%%%%%%%%%%%%%%%%%%%% T A B L E %%%%%%%%%%%%%%%%%%%%%%%%%%%%%%%%%%%%%%%%%%%%%%
\begin{table}
%\small
\caption{\label{tab:para-slep}
  MSSM default parameters for the numerical investigation; all 
  parameters (except of $\TB$) are in GeV.  The values for the trilinear 
  sfermion Higgs couplings, $A_f$ are chosen to be real (except for 
  $A_{\Fe_g}$ which can be complex) and such that charge- and/or 
  color-breaking minima are avoided \cite{ccb}.  
  It should be noted that we chose common values 
  $M_{\tilde Q, \tilde U, \tilde D} = 2000\gev$ for all squark generations,
  and $\MSL = \MSE + 50\gev$ for all slepton generations. 
}
\centering
\begin{tabular}{lrrrrrrrrrrrr}
\toprule
Scen. & $\sqrt{s}$ & $\TB$ & $\mu$ & $\MHp$ & $M_{\tilde Q, \tilde U, \tilde D}$ & 
$\MSE$ & $A_{\Fu_g}$ & $A_{\Fd_g}$ & $|A_{\Fe_g}|$ & $|M_1|$ & $M_2$ & $M_3$ \\ 
\midrule
\Scz & 1000 & 10 & 350 & 1200 & 2000 & 300 & 2600 & 2000 & 2000 & 400 & 
600 & 2000 \\
\bottomrule
\end{tabular}
\end{table}
%%%%%%%%%%%%%%%%%%%%% T A B L E %%%%%%%%%%%%%%%%%%%%%%%%%%%%%%%%%%%%%%%%%%%%%%

As in the previous subsection higher-order corrected Higgs-boson masses do not 
enter our calculation.
However, as before, we ensured that over larger parts of the parameter
space the lightest Higgs-boson mass is around $\sim 125 \pm 3\gev$ to
indicate the phenomenological validity of our scenarios.

%Fig7
%%%%%%%%%%%%%%%%%%%%%%%%%% F I G U R E %%%%%%%%%%%%%%%%%%%%%%%%%%%%%%%%%%%%%%%
\begin{figure}[t]
\begin{center}
\begin{tabular}{c}
\includegraphics[width=0.48\textwidth,height=6cm]{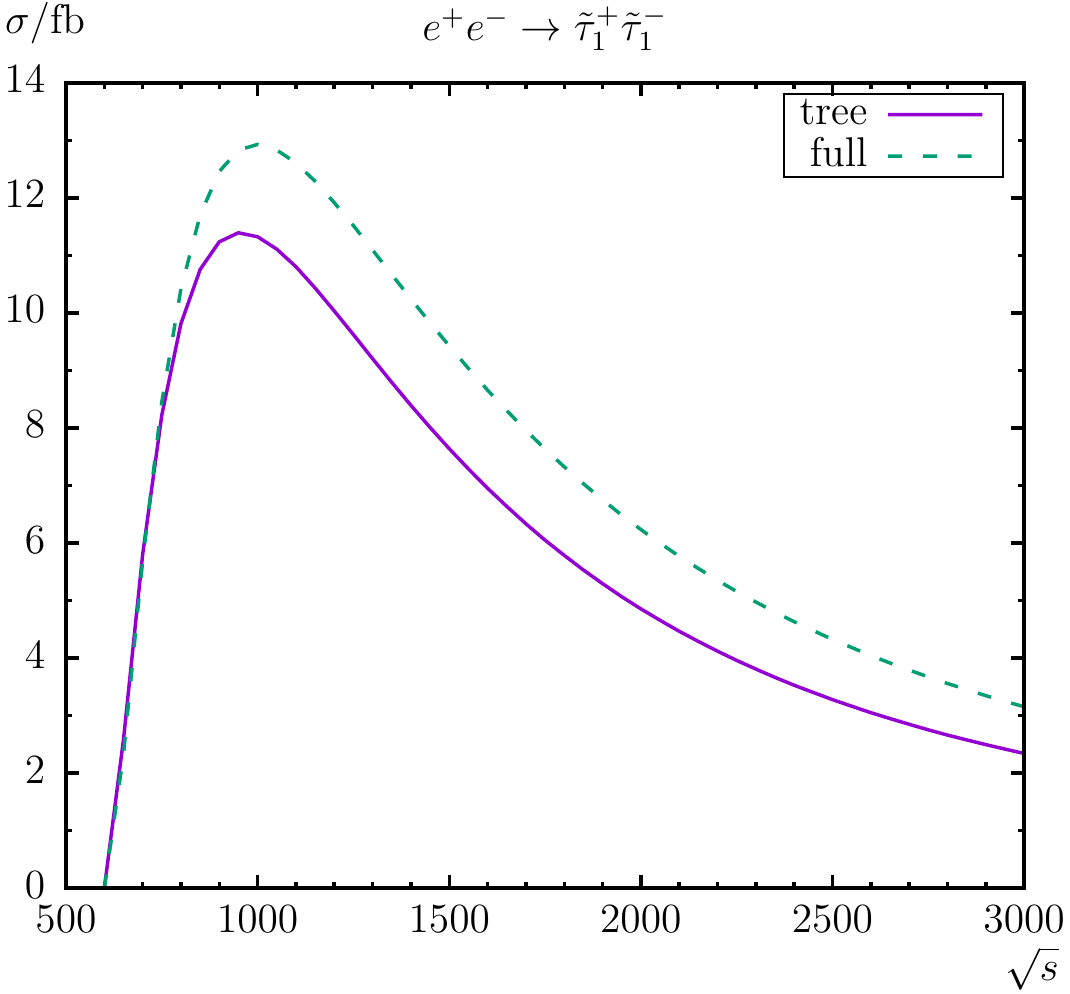}
\includegraphics[width=0.48\textwidth,height=6cm]{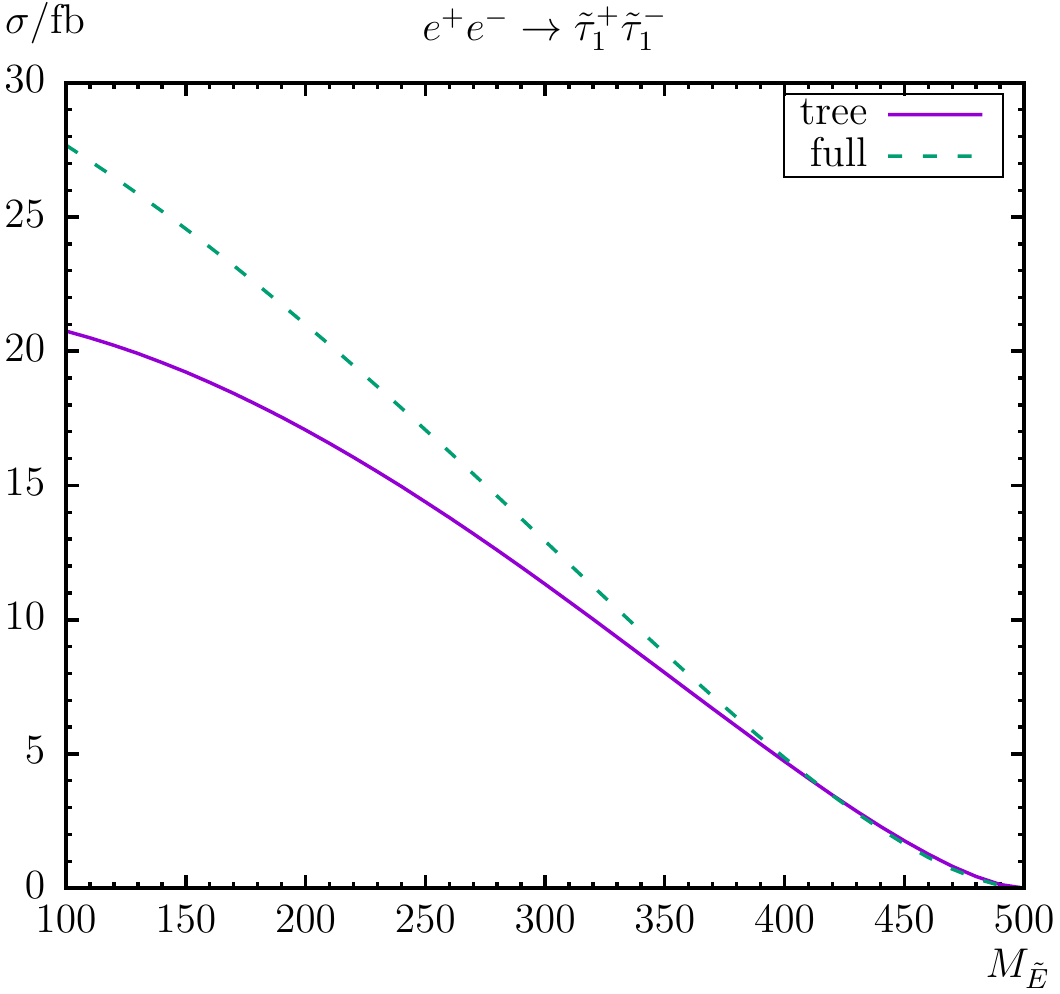}
\\[1em]
\includegraphics[width=0.48\textwidth,height=6cm]{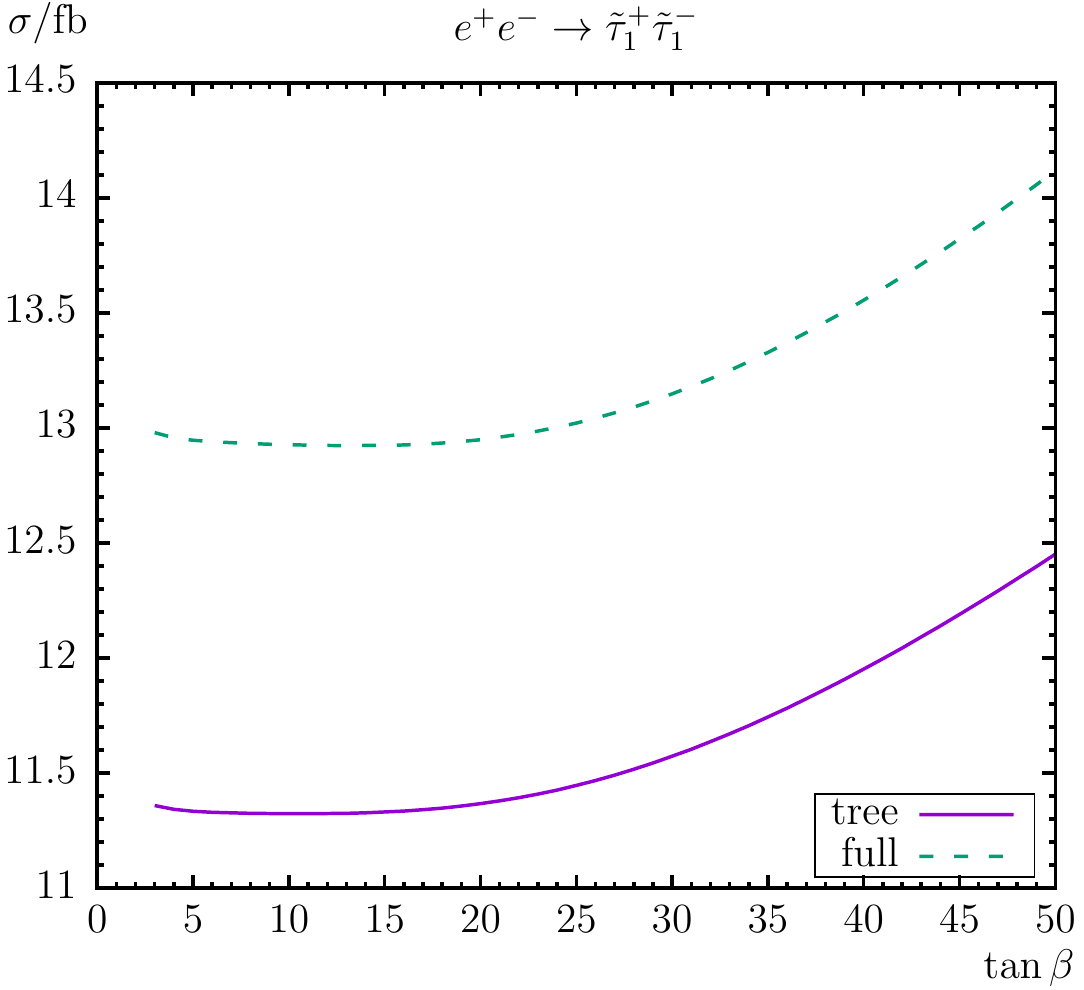}
\includegraphics[width=0.48\textwidth,height=6cm]{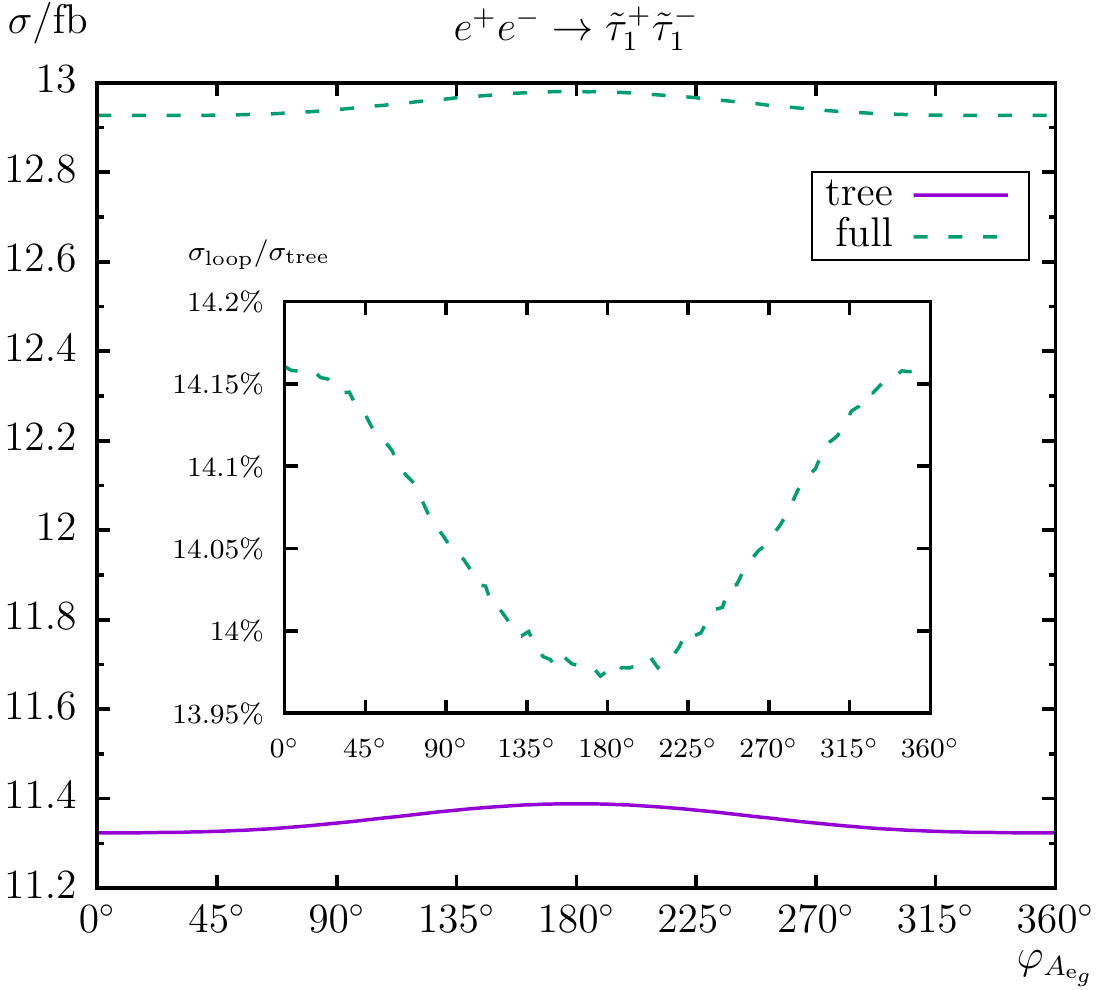}
\\%[1em]
\end{tabular}
\caption{\label{fig:eeSa1Sa1}
  $\sig(\eeSaeSae)$.
  Tree-level and full one-loop corrected cross sections are shown 
  with parameters chosen according to \Scz; see \refta{tab:para-slep}.
  The upper plots show the cross sections with $\sqrt{s}$ (left) 
  and $\MSE$ (right) varied; the lower plots show $\TB$ (left) and 
  $\phiAeg$ (right) varied. All masses and energies are in GeV.
}
\end{center}
\end{figure}
%%%%%%%%%%%%%%%%%%%%%%%%%% F I G U R E %%%%%%%%%%%%%%%%%%%%%%%%%%%%%%%%%%%%%%%

\bigskip
As an example of the numerical analysis that will be presented in
\citere{eeSlep} we show the 
process $\eeSaeSae$ in \reffi{fig:eeSa1Sa1}. 
As a function of $\sqrt{s}$ we find loop corrections of $\sim +14\,\%$ at 
$\sqrt{s} = 1000\gev$ (\ie \Scz), a tree crossing at 
$\sqrt{s} \approx 725\gev$ (where the one-loop corrections are
between $\pm 10\,\%$ for $\sqrt{s} \lsim 900 \gev$) and $\sim +35\,\%$ 
at $\sqrt{s} = 3000\gev$.

In the analysis as a function of $\MSE$ (upper right plot) the cross 
sections are decreasing with increasing $\MSE$ as obvious from 
kinematics and the full corrections have their maximum of $\sim 28\,\fb$ 
at $\MSE = 100\gev$, more than two times larger than in \Scz.  The relative 
corrections are changing from $\sim +33\,\%$ at $\MSE = 100\gev$ to 
$\sim -25\,\%$ at $\MSE = 490\gev$ with a tree crossing at $\MSE = 415\gev$.

In the lower left row of \reffi{fig:eeSa1Sa1} we show the dependence 
on $\TB$. The relative corrections for the $\TB$ dependence vary between 
$\sim +14.2\,\%$ at $\TB = 5$ and $\sim +13.4\,\%$ at $\TB = 50$.

The phase dependence $\phiAeg$ of the cross section in \Scz\ is shown 
in the lower right plot of \reffi{fig:eeSa1Sa1}.  
The loop correction increases the tree-level result by $\sim +14\,\%$.
The phase dependence of the relative loop correction is very small and 
found to be below $0.2\,\%$.
The variation with $\phiMe$ is negligible and therefore not shown here.

The production cross sections for the other sleptons will be published
in \citere{eeSlep}.

%%%%%%%%%%%%%%%%%%%%%%%%%%%%%%%%%%%%%%%%%%%%%%%%%%%%%%%%%%%%%%%%%%%%%%%%%%%%%%%
%%%%%%%%%%%%%%%%%%%%%%%%%%%%%%%%%%%%%%%%%%%%%%%%%%%%%%%%%%%%%%%%%%%%%%%%%%%%%%%

\section{Conclusions}
\label{sec:conclusions}

We have reviewed the calculation of chargino/neutralino/slepton
production modes at $e^+e^-$ colliders with a two-particle final state,
\ie \eecc, \eenn, \eeSeSe\ and \eeSnSn\ allowing for complex parameters, 
as given in \citeres{eeIno,eeSlep}. 
In the case of a discovery of charginos, neutralinos or sleptons a subsequent
precision measurement of their properties will be crucial to determine
their nature and the underlying (SUSY) parameters. 
In order to yield a sufficient accuracy, one-loop corrections to the 
various chargino/neutralino/slepton production modes have to be considered. 
This is particularly the case for the high anticipated accuracy of the
chargino/neutralino/slepton property determination at $e^+e^-$
colliders~\cite{LCreport}.

The evaluation of the processes (\ref{eq:eecc}) -- (\ref{eq:eeSnSn})
in \citeres{eeIno,eeSlep} 
is based on a full one-loop calculation, also including hard and soft 
QED radiation.  The renormalization is chosen to be identical as for 
the various chargino/neutralino/slepton decay calculations; see, \eg\
\citeres{Stau2decay,LHCxC,LHCxN,LHCxNprod} or chargino/neutralino/slepton
production from heavy Higgs boson decay; see, \eg\
\citeres{HiggsDecaySferm,HiggsDecayIno}.
Consequently, the predictions for the production and decay can be used 
together in a consistent manner (\eg\ in a global phenomenological 
analysis of the chargino/neutralino sector at the one-loop level).

For the analysis standard parameter sets (see
\reftas{tab:para-ccnn}, \ref{tab:para-slep}) were chosen, that allow the
production 
of all combinations of charginos/neutralinos or sleptons at an $e^+e^-$
collider with a center-of-mass energy up 
to $\sqrt{s} = 1000\gev$.

The review of the numerical analyses in \citeres{eeIno,eeSlep} showed
the following.
For the chargino pair production, \eecc, we observed an decreasing 
cross section $\propto 1/s$ for $s \to \infty$. 
The full one-loop corrections are very roughly 10-20\,\% of the
tree-level results, but depend strongly on the size of $\mu$, where
larger values result even in negative loop corrections.
The cross sections are largest for $\eecece$ and $\eeczcz$ and roughly
smaller by one order of magnitude for $\eececz$ due to the absence of the  
$\gamma\, \chapm{1} \champ{2}$ coupling at tree level in the MSSM.
The variation of the cross sections 
with $\phiMe$ or $\phiAt$ is found extremely small and the 
dependence on other phases were found to be roughly at the same level and 
have not been shown explicitely.

For the neutralino pair production, \eenn, the cross section can 
reach a level of \order{10\, \fb}, depending on the SUSY parameters,
but is very small for the production of two equal higgsino dominated
neutralinos at the \order{10\, \ab}.  This renders these processes 
difficult to observe at an $e^+e^-$ collider.%
\footnote{
  The limit of $10$~ab corresponds to ten events at an integrated 
  luminosity of $\cL = 1\, \iab$, which constitutes a guideline 
  for the observability of a process at a linear collider.
}
Having both beams polarized could turn out to be crucial to yield a
detectable production cross section in this case. 
The full one-loop corrections are very roughly 10-20\,\% of the tree-level 
results, but vary strongly on the size of $\mu$ and $\MSL$.
Depending on the size of in particular these two parameters the loop 
corrections can be either positive or negative.
The dependence on $\phiMe$ was found to reach up to $\sim 15\,\%$,
but can go up to $\sim 40\,\%$ for the extreme cases.  The (relative)
loop corrections varied by up to $5\,\%$ with~$\phiMe$. 

In the example shown for slepton production, $\sig(\eeSaeSae)$, the loop
corrections are found at the level of about $\sim \pm 15\%$, but 
vary strongly on the size of $\MSE$. While the phase and $\TB$ dependence 
is rather small, it can be larger for the other slepton production 
processes; see \citere{eeSlep}.

The given examples show that the loop corrections, including the complex
phase dependence, 
have to be included point-by-point in any precision analysis, or any 
precise determination of (SUSY) parameters from the production of
cMSSM charginos/neutralinos/sleptons at $e^+e^-$ linear colliders.

We emphasize again that our full one-loop calculation can readily be
used together with corresponding full one-loop corrections to
chargino/neutralino/slepton
decays~\cite{Stau2decay,LHCxC,LHCxN,LHCxNprod} or other 
chargino/neutralino/slepton production
modes~\cite{HiggsDecaySferm,HiggsDecayIno}.

%%%%%%%%%%%%%%%%%%%%%%%%%%%%%%%%%%%%%%%%%%%%%%%%%%%%%%%%%%%%%%%%%%%%%%%%%%%%%%

\subsection*{Acknowledgements}

The work of S.H.\ is supported 
in part by the MEINCOP Spain under contract FPA2016-78022-P, 
in part by the ``Spanish Agencia Estatal de Investigaci\'on'' (AEI) and the EU
``Fondo Europeo de Desarrollo Regional'' (FEDER) through the project
FPA2016-78022-P, 
and in part by
the AEI through the grant IFT Centro de Excelencia Severo Ochoa SEV-2016-0597.

%%%%%%%%%%%%%%%%%%%%%%%%%%%%%%%%%%%%%%%%%%%%%%%%%%%%%%%%%%%%%%%%%%%%%%%%%%%%%%%
%%%%%%%%%%%%%%%%%%%%%%%%%%%%%%%%%%%%%%%%%%%%%%%%%%%%%%%%%%%%%%%%%%%%%%%%%%%%%%%

%\newpage

\newcommand\jnl[1]{\textit{\frenchspacing #1}}
\newcommand\vol[1]{\textbf{#1}}

\end{document}